\documentclass[10pt, journal,compsoc]{IEEEtran}

\ifCLASSOPTIONcompsoc
  \usepackage[nocompress]{cite}
\else
  \usepackage{cite}
\fi

\usepackage{array,enumerate}
\usepackage{multirow}
\usepackage{graphicx}
\usepackage{booktabs}
\usepackage[linesnumbered,ruled,vlined]{algorithm2e}
\usepackage{float}
\SetArgSty{textup}
\usepackage{amssymb}
\usepackage{amsmath}
\usepackage{threeparttable}
\usepackage{bm}
\usepackage{algpseudocode}
\usepackage{ragged2e}
\usepackage{courier}
\usepackage{caption}
\usepackage{balance}

\newcolumntype{L}[1]{>{\raggedright\let\newline\\\arraybackslash\hspace{0pt}}m{#1}}
\newcolumntype{C}[1]{>{\centering\let\newline\\\arraybackslash\hspace{0pt}}m{#1}}
\newcolumntype{R}[1]{>{\raggedleft\let\newline\\\arraybackslash\hspace{0pt}}m{#1}}

\SetCommentSty{mycommfont}

\let\oldnl\nl
\newcommand{\nonl}{\renewcommand{\nl}{\let\nl\oldnl}}

\newcommand{\subcaption}[1]{\centerline{{\small
  #1}}\vspace{0pt}}
\newlength{\minipagewidth}
\newlength{\figurewidthFour}
\makeatletter
\newcommand{\thickhline}{%
    \noalign {\ifnum 0=`}\fi \hrule height 1.2pt
    \futurelet \reserved@a \@xhline
}
\newcolumntype{"}{@{\hskip\tabcolsep\vrule width 1pt\hskip\tabcolsep}}
\makeatother
\def\BibTeX{{\rm B\kern-.05em{\sc i\kern-.025em b}\kern-.08emT\kern-.1667em\lower.7ex\hbox{E}\kern-.125emX}}

\begin{document}

\title{\LARGE{\textbf{Optimizing Network Performance for Distributed DNN Training on GPU Clusters: ImageNet/AlexNet Training in 1.5 Minutes}}}

\author{Peng~Sun, Wansen~Feng, Ruobing~Han,  Shengen~Yan, and Yonggang~Wen 
    \IEEEcompsocitemizethanks{\IEEEcompsocthanksitem Peng~Sun,  Wansen~Feng, Ruobing~Han, and Shengen~Yan are with SenseTime. Email: \{sunpeng1, fengwansen, hanruobing, yanshengen\}@sensetime.com
    \IEEEcompsocthanksitem Yonggang Wen is with School of Computer Science and Engineering, Nanyang Technological University, Singapore. Email: ygwen@ntu.edu.sg}
}

\markboth{}%
{Shell \MakeLowercase{\textit{et al.}}: Bare Demo of IEEEtran.cls for Journals}

\IEEEtitleabstractindextext{

\begin{abstract}
\justifying
{It is important to scale out deep neural network (DNN) training for reducing model training time. The high communication overhead is one of the major performance bottlenecks for distributed DNN training across multiple GPUs. Our investigations have shown that popular open-source DNN systems could only achieve 2.5 speedup ratio on 64 GPUs connected by 56 Gbps network. To address this problem, we propose a communication backend named GradientFlow for distributed DNN training, and employ a set of network optimization techniques. First, we integrate ring-based allreduce, mixed-precision training, and computation/communication overlap into GradientFlow. Second, we propose lazy allreduce to improve network throughput by fusing multiple communication operations into a single one, and design coarse-grained sparse communication to reduce network traffic by only transmitting important gradient chunks. When training AlexNet and ResNet-50 on the ImageNet dataset using 512 GPUs, our approach could  achieve 410.2 and 434.1 speedup ratio respectively.}
\end{abstract}

\begin{IEEEkeywords}
Distributed Computing, Deep Learning, Computer Network
\end{IEEEkeywords}
}

\maketitle
\IEEEdisplaynontitleabstractindextext
\IEEEpeerreviewmaketitle

\section{Introduction}

Deep neural network (DNN) builds models from training data, and uses them to make predictions on new data. It has been  used in a wide range of applications, including image recognition, natural language processing and recommender systems. Typically, a DNN model has tens or hundreds of layers, each of which consists of a large number of parameters represented by tensors. To minimize the prediction error, a DNN training task usually uses an iterative-convergent algorithm to iteratively compute gradients from training datasets, and aggregate them with the current version of parameters.

With the ever-increasing sizes of training datasets and bigger models, DNNs are getting more computationally expensive to train on a single node. For example, completing 90-epoch ResNet-50 \cite{he2015deep} training on the ImageNet-1K \cite{russakovsky2014imagenet} dataset takes 14 days using a single M40 GPU, and takes 29 hours using a machine with 8 NVLink-connected P100 GPUs \cite{goyal2017accurate}. This is true even when leveraging high-performance DNN computation libraries like cuDNN \cite{chetlur2014cudnn}, which could achieve near the theoretical peak computation performance on GPUs in many cases. Moreover, users often run multiple training tasks to achieve the best result for a specific mission \cite{Gandiva}. In this case, extremely long model training time significantly impedes the research and development progress. It has been shown that training time is a key challenge at the root of the development of new DNN architectures \cite{43150}.

\setlength{\minipagewidth}{0.235\textwidth}
\setlength{\figurewidthFour}{\minipagewidth}
\begin{figure}
    \centering
    \begin{minipage}[t]{\minipagewidth}
    \begin{center}
    \includegraphics[width=\figurewidthFour]{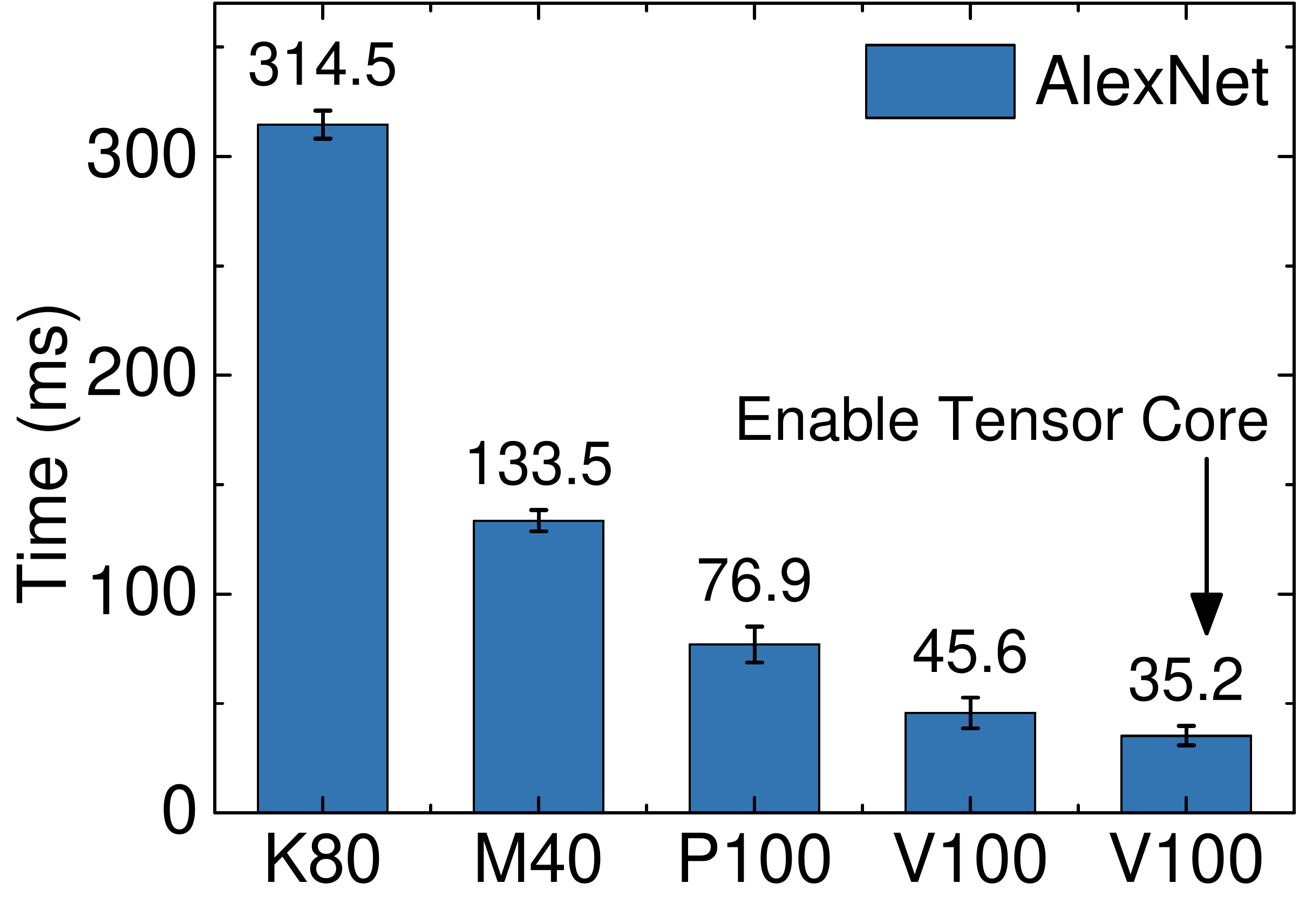}
    \subcaption{(a) AlexNet}
    \end{center}
    \end{minipage}
    \centering
    \begin{minipage}[t]{\minipagewidth}
    \begin{center}
    \includegraphics[width=\figurewidthFour]{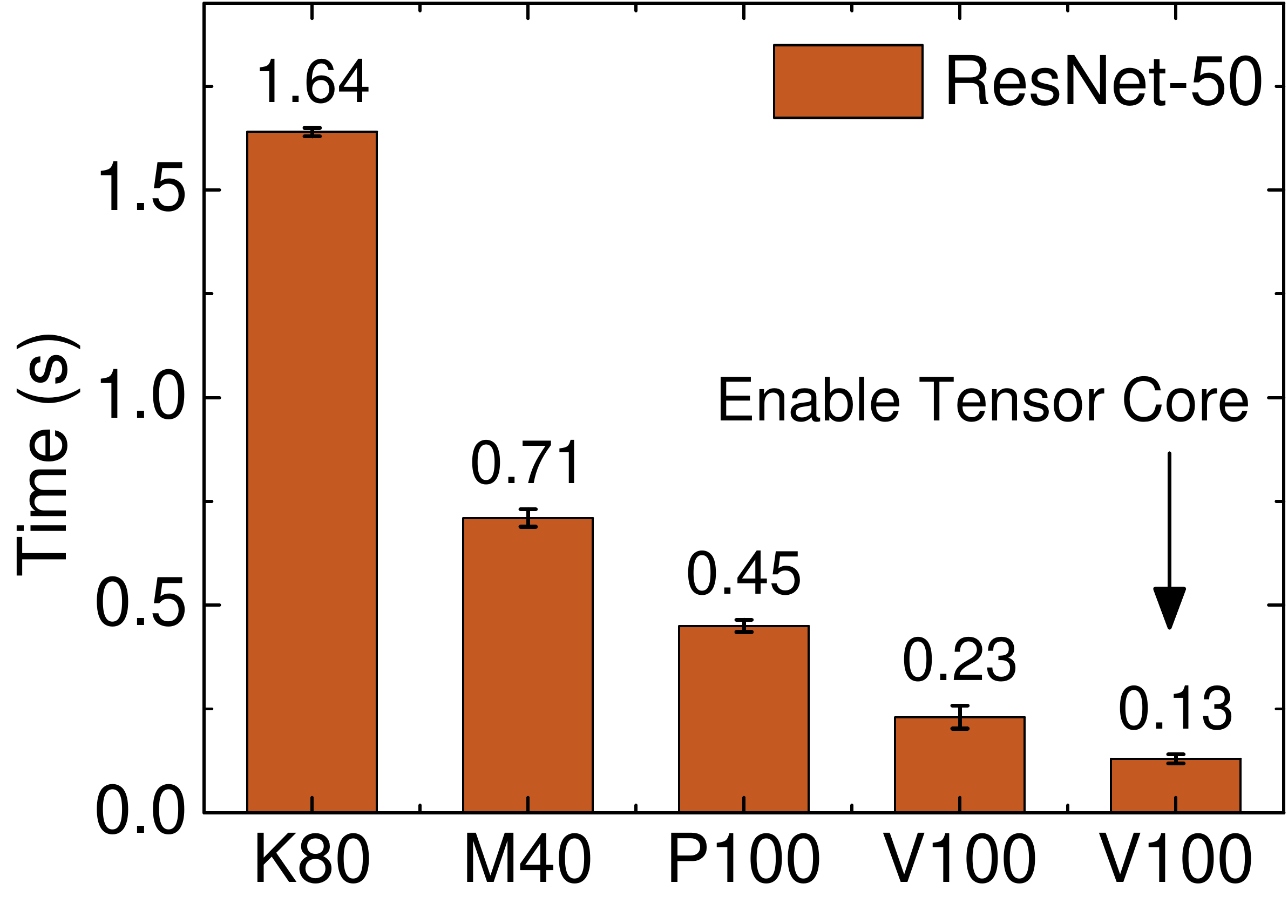}
    \subcaption{(b) ResNet-50}
    \end{center}
    \end{minipage}
    \centering
    \caption{Single-GPU per-iteration training time for AlexNet (batch size 128) and ResNet-50 (batch size 64) measured with CuDNN-7 on  K80, M40, P100 and V100 GPUs, which were launched on 2014, 2015, 2016 and 2017, and use Kepler, Maxwell, Pascal and Volta micro-architectures, respectively.}
\label{Fig: BaseSinglePerf}
\end{figure}

Various distributed DNN systems have been proposed to accelerate model training, such as MxNet \cite{chen2015mxnet}, PyTorch \cite{Pytorch},  TensorFlow \cite{199317} and Petuum \cite{Petuum}, Adam \cite{chilimbi2014project} and GeePS \cite{cui2016geeps}. These systems usually adopt data parallelism to parallelize model training across multiple GPUs or multiple machines. In this method, the training dataset is split into $N$ parts stored on each GPU. During model training, each GPU performs computation on a batch of allocated training dataset, and leverages parameter server \cite{li2014scaling} or allreduce \cite{sparks2013mli} to exchange its generated gradients with other GPUs via network for updating the DNN model's parameters. This is done iteratively to bring all layers' parameters closer to the optimal values. Communication backend is a key component of distributed DNN systems: it uses network to connect multiple GPUs, and decides how GPUs exchanges intermediate data during model training.

A major performance bottleneck of large-scale distributed DNN training is the high communication overhead due to following  factors. First, the computation power of GPU accelerators dramatically increases in recent years, causing bandwidth deficiency. Figure \ref{Fig: BaseSinglePerf} shows that the single-GPU per-iteration processing time of two classic DNNs, AlexNet \cite{krizhevsky2012imagenet} and ResNet-50 \cite{he2015deep}, has decreased by 8.9x and 12.6x from 2014 to 2017, respectively. The increasing computation power demands a similar increase in network bandwidth to avoid communication bottleneck. However, upgrading datacenter networks is expensive: network bandwidth on major cloud datacenters and HPC clusters has improved little across generational upgrades \cite{ParameterHub}. For example, the first HPC cluster built for DNN training incorporates 56Gbps network in 2013 \cite{coates2013deep}, and a recent one uses 100Gbps network in 2018 \cite{jia2018highly}. Second, DNNs are trending to learn large models with tens or hundreds of millions of parameters, generating a large amount of network traffic for distributed training. Considering the increasing GPU computation power, distributed DNN systems may spend a significant portion of time on communication. For example, when training AlexNet with $61$M single-precision parameters, the total amount of data transferred through every GPU is $488$MB in each iteration. Even if the communication backend  maximizes the throughput of 56Gbps network, each GPU still takes $69.7$ms for communication, which could be even longer than the per-iteration computation time (see Figure \ref{Fig: BaseSinglePerf}). Third, the communication strategies used in popular distributed DNN systems could not fully utilize cluster network resources. When training AlexNet on a 64 GPU-cluster connected by 56Gbps network, our experiments show that PyTorch, which uses Gloo as its default communication backend, roughly takes $2150$ms for communication per iteration, and only utilizes $3.3\%$ of the cluster's available network bandwidth. Hence, it is important to reduce the communication overhead for training DNNs on large-scale GPU clusters.

A number of approaches have been proposed to reduce the communication cost for distributed DNN training. Compared to commodity network (e.g., 1/10Gbps network), high-performance network fabrics like InfiniBand could provide much higher bandwidth (e.g., 56/100 Gbps) for GPUs to exchange gradients  \cite{coates2013deep}, \cite{wu2015deep}, \cite{li2015malt}, \cite{iandola2016firecaffe}. Remote Direct Memory Access (RDMA), GPUDirect RDMA and CUDA-aware MPI further reduce GPU-GPU communication latency by enabling direct data exchange between GPUs on different nodes  \cite{awan2017s}, \cite{wang2011mvapich2}. When exchanging gradients with ring-based allreduce, the communication cost is constant and independent of the number of GPUs in the system \cite{Baidu-AllReduce}, \cite{NCCL}. Model compression approaches represent parameters and gradients with limited numerical precision or sufficient factors, thus reducing data size for transmission \cite{jia2018highly}, \cite{seide2014parallelizability}, \cite{seide20141}, \cite{gupta2015deep}, \cite{micikevicius2017mixed}, \cite{zhang2015poseidon}. Asynchronous and stale-synchronous parallelisms do not force all GPUs to wait for the slowest one in every iteration, thus reducing the synchronization time  in heterogeneous environments  \cite{ho2013more}, \cite{recht2011hogwild}.  Overlapping the communication of the DNN's top layers with the computation of bottom layers could also help to reduce communication time for distributed DNN training \cite{awan2017s}, \cite{zhang2015poseidon}, \cite{hashemi2018tictac}, \cite{sergeev2018horovod}. Some distributed DNN systems support to select and transmit important gradients at the level of elements  \cite{li2014scaling}, \cite{sun2016timed}, \cite{lin2017deep}, \cite{D17-1045}, reducing network traffic.

The relative performance characteristics of aforementioned approaches are still unclear. Although each one reports performance results, they are not easily comparable, since their experiments are conducted with different models and on different infrastructures (usually with less than 100 GPUs). Our first goal is to address this issue. Specifically, we implement GradientFlow, a communication library for distributed DNN, with multiple network optimizations (including ring-based allreduce, mixed-precision training, layer-based communication/computation overlap and element-level sparse communication), and integrates it with our internal distributed DNN system. To evaluate the performance, we measure processing throughput  to train two classic DNNs, AlexNet and ResNet-50, on the ImageNet-1K dataset using two physical GPU clusters with 56 Gbps network.

Based on the benchmarking results, communication is still a big challenge for distributed DNN. Our second goal is to tackle this problem with following two techniques in GradientFlow:
1) To improve network throughput, we propose lazy allreduce for gradient exchanging. Instead of immediately transmitting generated gradients with allreduce, GradientFlow tries to fuse multiple sequential communication operations into a single one, avoiding sending a huge number of small tensors via  network.
2) To reduce network traffic, we design coarse-grained sparse communication. Instead of transmitting all gradients in every iteration, GradientFlow only sends important gradients for allreduce at the level of chunk (for example, a chunk may consist of  32K gradients).  GradientFlow imposes momentum SGD correction and warm-up dense training to guarantee model quality. Compared to existing fine-grained sparse communication strategies, e.g.,  \cite{li2014scaling}, \cite{sun2016timed}, \cite{lin2017deep},  coarse-grained sparse communication could keep high bandwidth utilization by using allreduce with dense inputs.

When training  ImageNet/AlexNet, our internal distributed DNN system could achieve  410.2 speedup ratio on 512 GPUs, and complete 95-epoch training in 1.5 minutes. Compared to Jia et al. \cite{jia2018highly}, which finishes 95-epoch ImageNet/AlexNet training in 4 minutes with 1024 GPUs, our work could achieve 2.6x speedup. When training ImageNet/ResNet-50, our approach could achieve 434.1  speedup ratio on 512 GPUs, and complete 90-epoch training in 7.3 minutes. Compared to  Akiba et al. \cite{akiba2017extremely}, which finishes ResNet-50 training in 15 minutes with 1024 GPUs, our work is 2.1x faster.

The rest of this paper is structured as follows. We show the background of distributed DNN, and introduce existing network optimizations for distributed DNN with performance evaluations. in Section 2.  Section 3 describes system design of lazy allreduce and coarse-grained sparse communication.  The overall evaluation results are detailed in Section 4.  Section 5 concludes this paper.

\section{Distributed DNN Training}

In this section, we briefly summarize basic concepts of distributed DNN training. Then, we describe the baseline implementation of a distributed DNN training system.  Finally, we show existing network optimizations for distributed DNN training and evaluate their performance.

\subsection{Background of DNN}

DNNs learn models from training data, and use them to make predictions on new data.  Typically, a DNN consists of many layers, from as few as 5 to as many as more than 1000 \cite{lecun2015deep}.  Figure \ref{Fig: DNN_arch} shows a DNN with 7 layers. In this example, the data layer is in charge of reading and preprocessing input data. The first layer is connected to a sequence of intermediate layers (two convolutional layers, two pooling layers, one innerproduct layer and one loss layer), each of which uses a specific function and a number of parameters to transform its input to an output. Finally, the DNN  transforms the raw input data into the desired output or prediction.

\setlength{\minipagewidth}{0.475\textwidth}
\setlength{\figurewidthFour}{\minipagewidth}
\begin{figure}[H]
    \centering
    \includegraphics[width=\figurewidthFour]{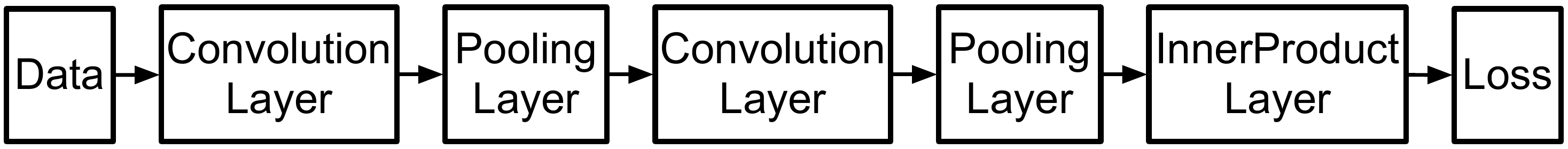}
    \caption{A deep neural network with 7 layers.}
\label{Fig: DNN_arch}
\end{figure}

To give accurate predictions, most DNNs need to be trained.  We use  $W$ to represent all parameters  of a DNN. DNN training could be described  as an optimization problem: given training dataset $\mathcal{D}$ with $m$ examples, it tries to find optimum $W$ to minimize the objective function $J(\cdot)$,
$$
\underset{W} \min \quad J(W) = \sum\nolimits_{i=1}^{m}{f(W, \mathcal{D}_i)} + r{(W)},
$$
where $f(\cdot)$ is the loss function to represent the DNN's prediction error on one example of training dataset, and $r(\cdot)$ is the regularizer to limit the learned DNN model's complexity.

It is common to use an iterative-convergent algorithm and backpropagation (BP) to train DNNs. Each iteration has three sequential phases. 1) The forward pass transforms a batch of input examples into predictions, which would be compared with given labels to calculate prediction error. 2) With respect to this error, the backward pass calculates gradients for each layer's learnable parameters through BP. 3) In the model update stage, each parameter is updated using its corresponding gradient and a variant of the gradient descent optimization algorithm, such as stochastic gradient descent (SGD).

\subsection{Data-Parallel Distributed DNN Training}

It is hard to finish model training with big training datasets in acceptable time on a single node. To address this issue, many distributed DNN systems, such as TensorFlow and PyTorch, have been proposed to parallel DNN training in a cluster using data-parallel  strategy\footnote{Model-parallelism is another approach, which is designed to deal with DNN applications with extreme-big models. In this work, we focus on data parallel DNN training.}, where the same model is replicated for every GPU, but is fed with different parts of training data.

\setlength{\minipagewidth}{0.475\textwidth}
\setlength{\figurewidthFour}{\minipagewidth}
\begin{figure}
    \centering
    \includegraphics[width=\figurewidthFour]{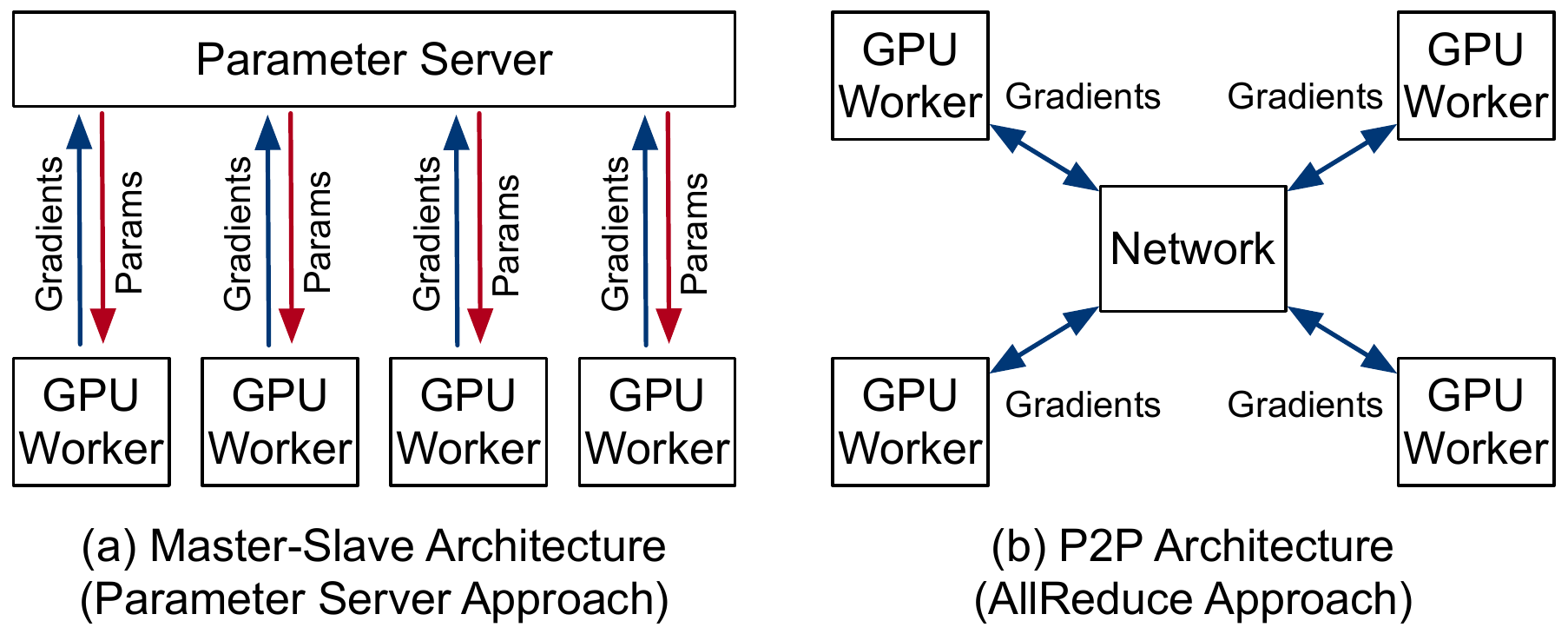}
    \caption{System architectures for distributed DNN. In master-slave architecture, all GPU workers push local gradients to parameter servers for model update, and pull latest parameters. In P2P architecture, every GPU worker  communicates with each other to exchange gradients using allreduce.}
\label{Fig: ps_allreduce}
\end{figure}

As shown in Figure \ref{Fig: ps_allreduce}, there are two design choices to implement data-parallel DNN training: the parameter server (PS) approach using master-slave architecture and the allreduce approach with P2P architecture. In PS, one or more \emph{server} nodes are set up to centrally manage the model's parameters. For every iteration, each worker pushes its computed gradients to server nodes for aggregation and model update, and pulls latest parameters from server nodes. In the allreduce approach, all workers directly communicate with each other to exchange local gradients using allreduce operations. After the allreduce operation, every GPU has aggregated gradients, and uses them to update replicated parameters locally.

In this paper, we have three  assumptions for distributed DNN training: 1) a single GPU has enough storage capacity to manage a complete replica of the model; 2) each layer's parameters and its corresponding gradients could be represented by one or more dense tensors; and 3) training tasks are deployed in a homogeneous cluster: all computation machines are installed with same type of GPUs and network devices.
Considering aforementioned assumptions, we focus on network optimization for allreduce-based distributed DNN training. While PS provides flexible synchronization mechanisms, supports extra-big model training, and can  provide better scalability than some allreduce-based approaches \cite{BytePS}, it requires a heterogeneous cluster to provide separated parameter server machines.  Although we could use half of GPU machines to serve as CPU-based parameter server nodes, this methods would waste a huge amount of computation power in our homogeneous clusters.

\subsection{Baseline System Design \& Cluster Setting}

Figure \ref{Fig: distributed_dnn_baseline} illustrates the baseline system design for allreduce-based distributed DNN training. In the forward pass, each GPU fetches a batch of training data as input, and processes them through the neural network from layer-$1$ to layer-$n$. Assuming each GPU processes $B$ images per iteration, the training task's batch size is $NB$ with $N$ GPUs. At the end of the forward pass, each GPU outputs a loss value to measure the prediction error. Next, GPUs start backward computation through the neural work from layer-$n$ to layer-$1$. When completing the backward computation of layer-$i$, this layer's gradients are generated for its learnable parameters. Noted that not all layers have learnable parameters, such as ReLu layer and pooling Layer. Some layers may use multiple tensors to represent its parameters and gradients. For instance, convolutional layer writes a \emph{weight} gradient tensor and a \emph{bias} gradient tensor during backward computation. After the backward pass, a set of gradient tensors are generated on every GPU. An allreduce operation is applied to each tensor for gradient aggregation. After the communication stage, each GPU has summed gradients, which could be used to update replicated parameters locally. By default, all parameter and gradient tensors are composed of single-precision (FP32) values in the baseline system.

\setlength{\minipagewidth}{0.475\textwidth}
\setlength{\figurewidthFour}{\minipagewidth}
\begin{figure}
    \centering
    \includegraphics[width=\figurewidthFour]{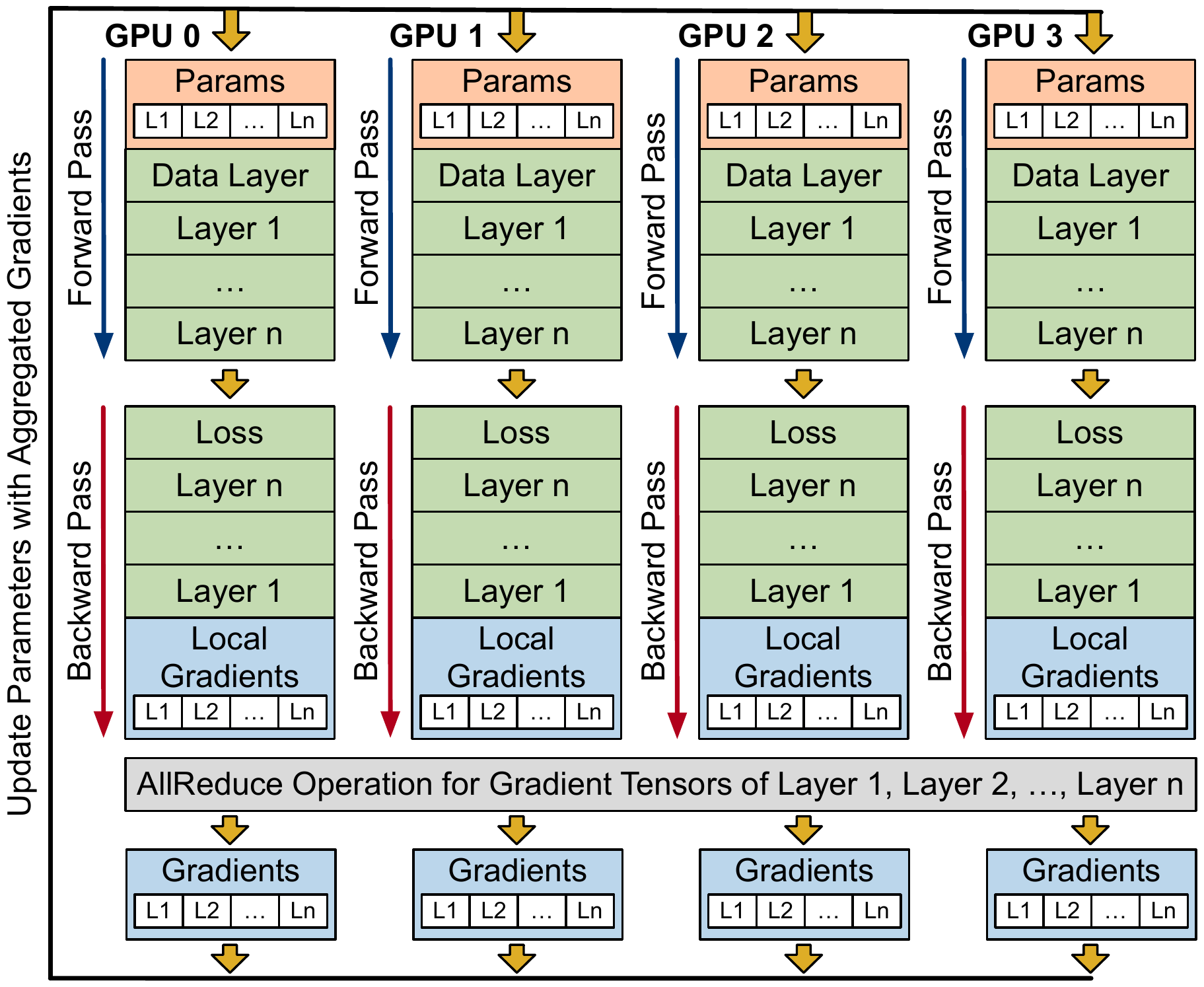}
    \caption{Baseline system for distributed DNN training.}
\label{Fig: distributed_dnn_baseline}
\end{figure}

% \textbf{\emph{Benchmarked DNNs.}}
\subsubsection{Benchmarked DNNs}

To evaluate system performance, we measure the training time of two classic DNNs, AlexNet and ResNet-50, on the ImageNet-1K dataset, which contains more than $1.2$M labeled images and 1000 classes. The neural network structure of AlexNet used in this paper is slightly different with the origin one in \cite{krizhevsky2012imagenet}. To achieve better model quality, we use the version of AlexNet in \cite{jia2018highly} with proper batch normalization layers. Note that we do not consider synchronized batch normalization in this work. Figure \ref{Fig: DNN-info} shows the information of AlexNet and ResNet-50. Specifically, AlexNet contains 27 layers with $60.9$M learnable parameters. The corresponding values for ResNet-50 are 188 and $25.5$M. Since the training system should output a gradient tensor for each learnable parameter tensor, the backward computation generates 26 and 153 gradient tensors for AlexNet and ResNet-50, respectively. In Figure \ref{Fig: DNN-info}, we label  learnable parameter tensors with ID in ascending order from layer-$1$ to layer-$n$. Since the backward pass performs computation from layer-$n$ to layer-$1$, gradient tensors are generated in descending order by ID.

\setlength{\minipagewidth}{0.475\textwidth}
\setlength{\figurewidthFour}{\minipagewidth}
\begin{figure}
    \centering
    \begin{minipage}[t]{\minipagewidth}
    \begin{center}
    \includegraphics[width=\figurewidthFour]{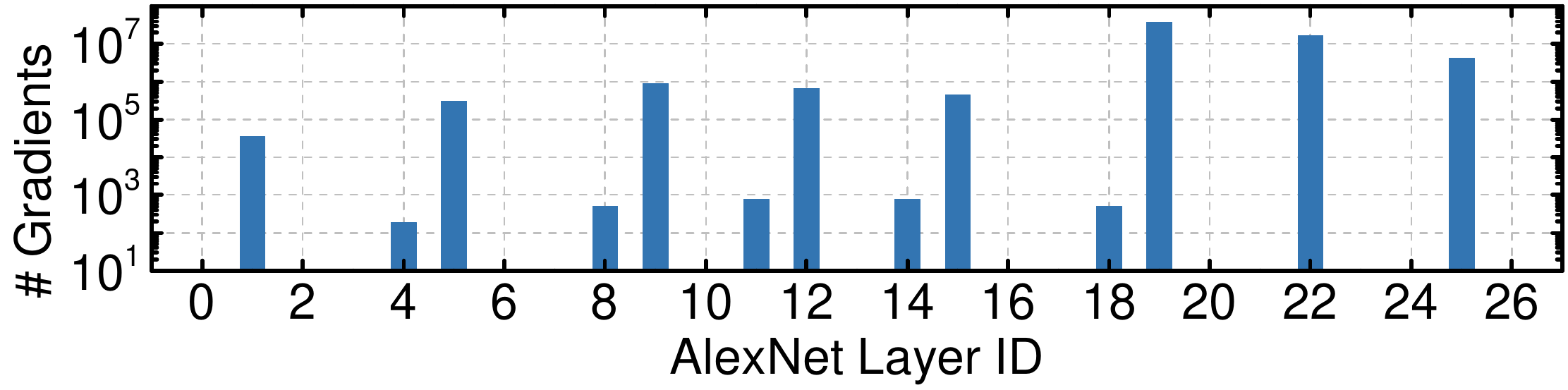}
    \end{center}
    \end{minipage}
    \centering
    \\[5pt]
    \centering
    \begin{minipage}[t]{\minipagewidth}
    \begin{center}
    \includegraphics[width=\figurewidthFour]{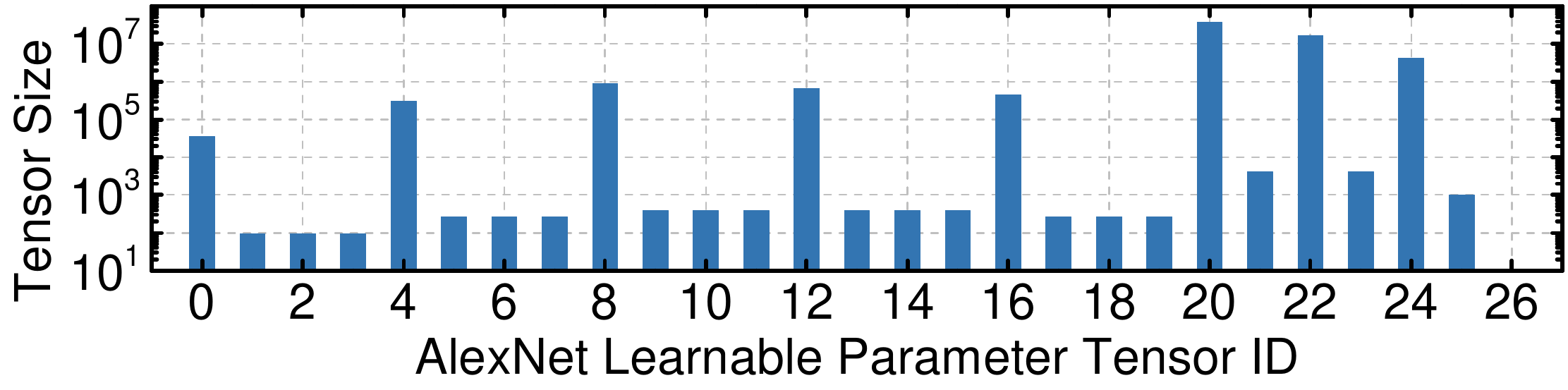}
    \end{center}
    \end{minipage}
    \centering
    \\[5pt]
    \begin{minipage}[t]{\minipagewidth}
    \begin{center}
    \includegraphics[width=\figurewidthFour]{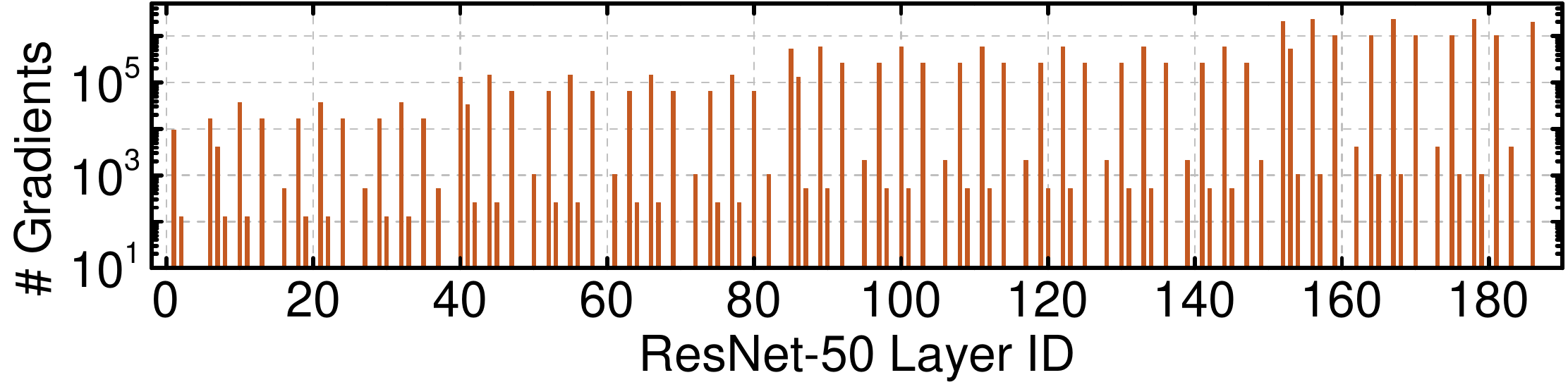}
    \end{center}
    \end{minipage}
    \centering
    \\[5pt]
    \begin{minipage}[t]{\minipagewidth}
    \begin{center}
    \includegraphics[width=\figurewidthFour]{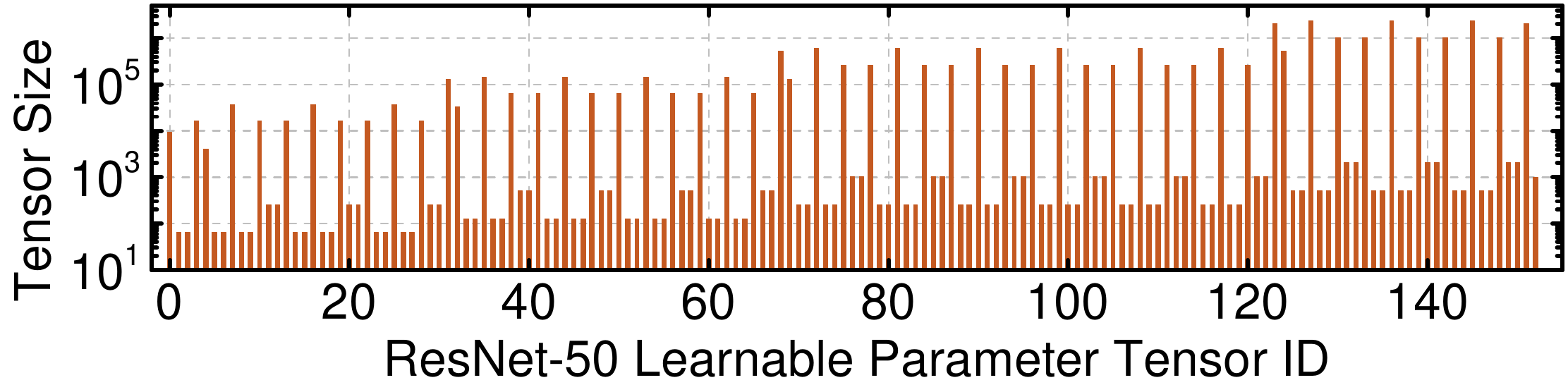}
    \end{center}
    \end{minipage}
    \centering
\caption{Information of AlexNet and ResNet-50. AlexNet has 27 layers with $60.9$M learnable parameters managed by 26 tensors. ResNet-50 has 188 layers with $25.5$M learnable parameters managed by 152 tensors.}
\label{Fig: DNN-info}
\end{figure}

\subsubsection{Cluster Hardware \& Software Setting}
% \textbf{\emph{Cluster Hardware \& Software Setting.}}

We use two clusters for performance measurement: Cluster-P and Cluster-V.
\begin{itemize}
\item Cluster-P contains 16 physical machines and 128 GPUs with Pascal microarchitecture.
\item Cluster-V contains 64 physical machines and 512 GPUs with Volta microarchitecture. Compared to Pascal GPUs, Volta GPUs could use Tensor Cores to accelerate matrix operations by performing mixed-precision matrix multiply and accumulate calculations in a single operation.
\end{itemize}
In both clusters, each physical machine is installed with 8 GPUs using PCIe-switch architecture: 4 GPUs under one PCIe switch, and each machine contains two PCIe switches. All machines of a cluster are connected by 56Gbps InfiniBand, and share a distributed file system, which is used for training dataset management. In this paper, two distributed DNN training systems are used to benchmark different communication strategies: PyTorch-0.4 and our internal system (System-I). Both systems are compiled with Cuda-9.0 and CuDNN-7.2, and use SGD for model training.

% \textbf{\emph{Baseline System Performance Evaluation. }}
\subsubsection{Baseline System Performance Evaluation}
We scale out PyTorch and System-I based on the architecture shown in Figure \ref{Fig: distributed_dnn_baseline}. In this set of experiments, we use Gloo, the default communication backend of PyTorch, to perform allreduce operations in PyTorch, and use OpenMPI-3.1 in System-I. Both Gloo and OpenMPI support to use RDMA for inter-node communication. In our baseline system with OpenMPI, we use butterfly algorithm for allreduce operations. Each GPU processes 128 and 64 images at each iteration for AlexNet and ResNet-50, respectively. The two baseline systems perform single-precision training on Cluster-P. Figure \ref{Fig: Baseline_Pytorch_Sensenet} shows the performance evaluation results.

\setlength{\minipagewidth}{0.235\textwidth}
\setlength{\figurewidthFour}{\minipagewidth}
\begin{figure}
    \centering
    \begin{minipage}[t]{\minipagewidth}
    \begin{center}
    \includegraphics[width=\figurewidthFour]{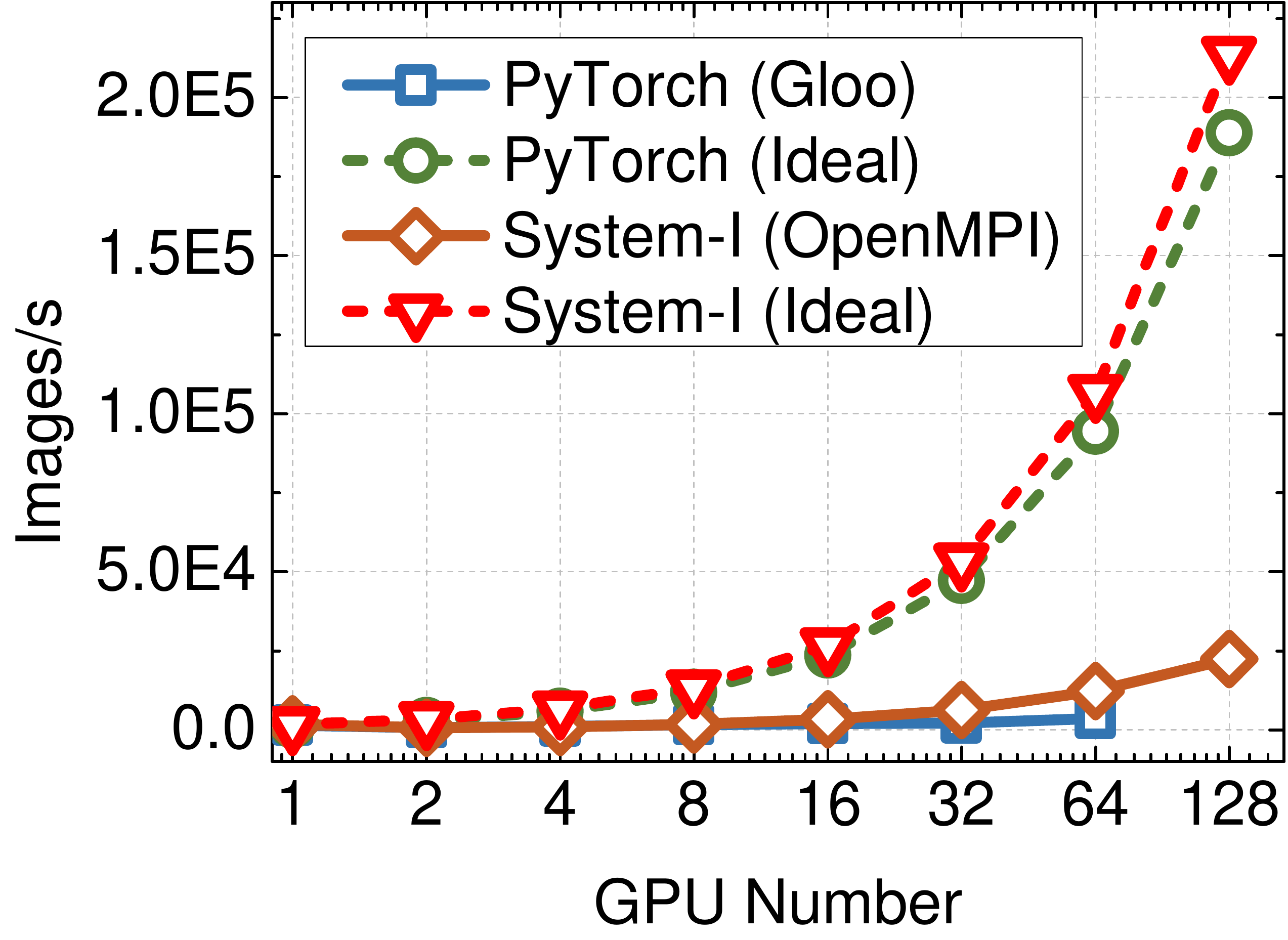}
    \subcaption{(a) AlexNet}
    \end{center}
    \end{minipage}
    \centering
    \begin{minipage}[t]{\minipagewidth}
    \begin{center}
    \includegraphics[width=\figurewidthFour]{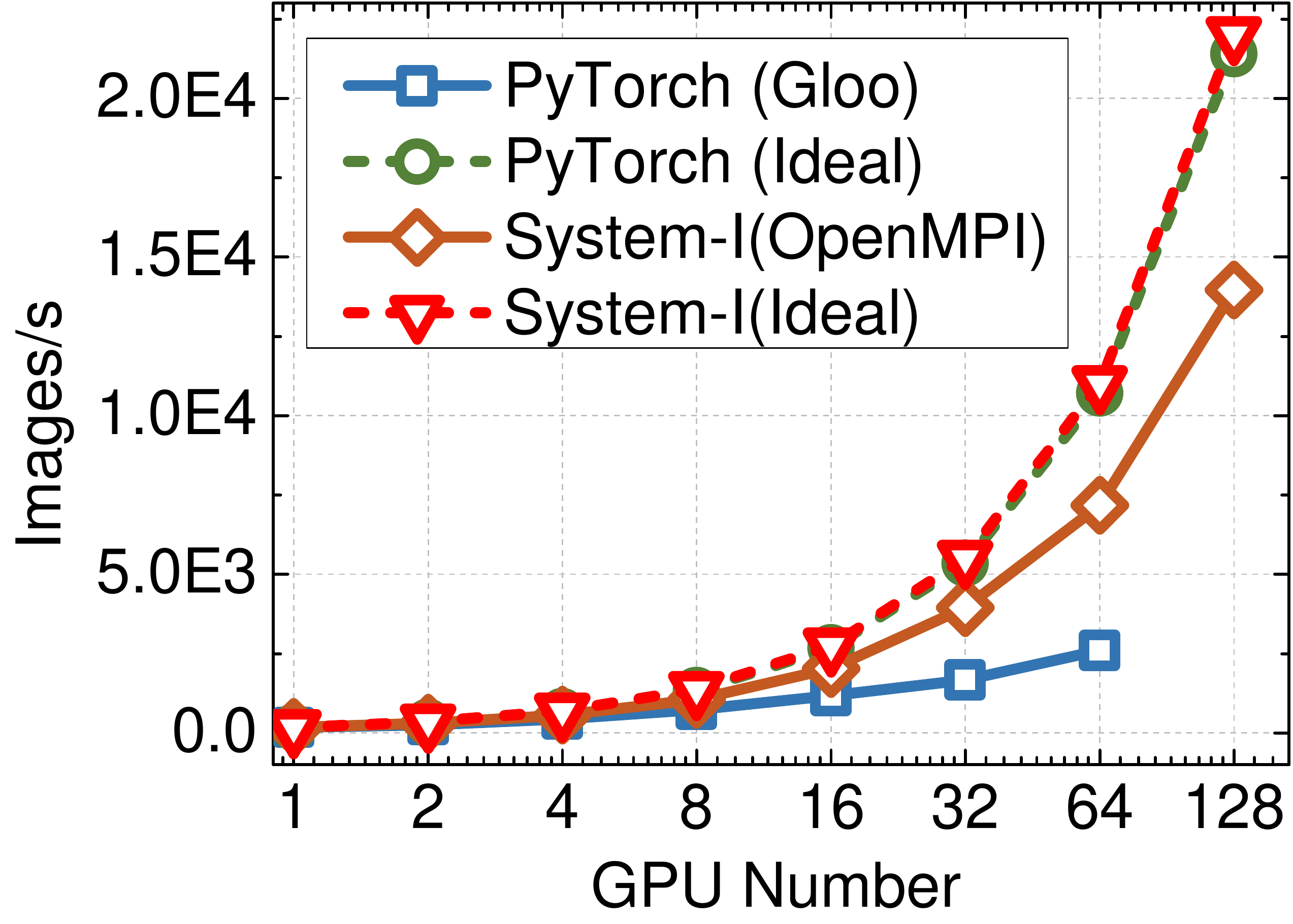}
    \subcaption{(b) ResNet-50}
    \end{center}
    \end{minipage}
    \centering
\caption{Baseline system (single-precision) performance evaluation in Cluster-P. The per-GPU batch size is 128 and 64 for AlexNet and ResNet-50 respectively.}
\label{Fig: Baseline_Pytorch_Sensenet}
\end{figure}

PyTorch and System-I have similar processing throughput when training AlexNet and ResNet-50 on a single GPU. Specifically, PyTorch and System-I could respectively process 1475 and 1670 images per second for AlexNet. The corresponding values for ResNet-50 are 167 and 172.

Due to the high communication cost,  PyTorch and System-I with the baseline design have poor scalability. When training AlexNet, PyTorch (Gloo)  could only process 3.6K images per second using 64 GPUs (with speedup ratio 2.5), and fails to run on 128 GPUs. While System-I (MPI) has better scalability than PyTorch (Gloo), the throughput is 22.3K image/s  with 128 GPUs (with speedup ratio 13.1). Since ResNet-50 has less learnable parameters and takes longer forward/backward computation time than AlexNet (see Figure \ref{Fig: DNN-info}), its communication overhead is not as significant as AlexNet. However, there is still a huge performance gap between the ideal system with two baseline implementations. Specifically, the speedup ratio is 15.5  on 64 GPUs with PyTorch (Gloo), and is 81.2  on 128 GPUs with System-I (MPI), when training ResNet-50.

\subsection{Ring-based AllReduce}

\setlength{\minipagewidth}{0.47\textwidth}
\setlength{\figurewidthFour}{\minipagewidth}
\begin{figure} 
    \includegraphics[width=\figurewidthFour]{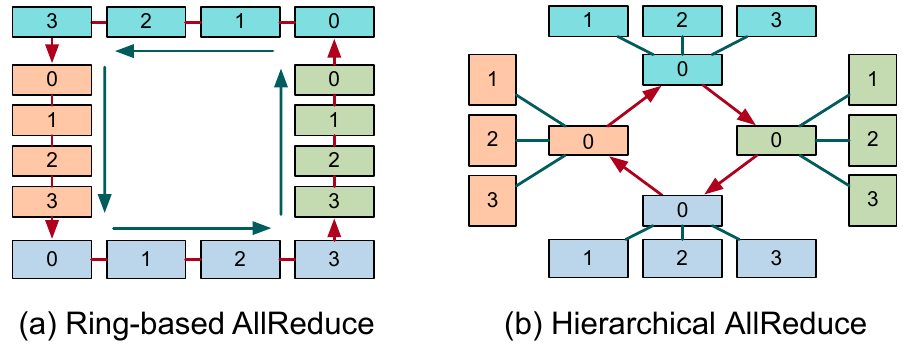}
    \centering
    \caption{Ring-based allreduce and hierarchical allreduce. In this example, there are 4 machines, each of which has 4 GPUs. In (a), all 16 GPUs are arranged in a logical ring. In (b), the 16 GPUs are partitioned into 4 groups, and only the master GPU of each group constructs a logical ring for allreduce.}
\label{Fig: arch_allreduce}
\end{figure}

An efficient allreduce algorithm and implementation is vital for distributed DNN. Ring-based allreduce \cite{Baidu-AllReduce} is an algorithm to perform allreduce with constant communication cost, which is measured by the amount of data transferred to and from every machine. As shown in Figure \ref{Fig: arch_allreduce}(a), all $N$  GPUs are arranged in a logical directed ring, and the $K$ bytes of input tensor are equally partitioned into $N$ chunks. Each GPU sends and receives $K/N$ bytes of data $2(N-1)$ times to complete an allreduce operation. Thus, the total amount of network traffic through each GPU is $2(N-1)K/N$, which is independent of the number of GPUs. In large-scale clusters, the ring-based allreduce algorithm may not fully utilize bandwidth due to small message size. Hierarchical allreduce \cite{jia2018highly} is proposed to address this issue. As shown in Figure \ref{Fig: arch_allreduce}(b), this method groups $N$ GPUs into $M$ groups, and uses three phases to do allreduce: 1) each group does a ring-based reduce to store  partial results to its master GPU, 2) a ring-based allreduce is launched on $M$ master GPUs, after each master GPU gets the final result, and 3) each master GPU does a broadcast within each group to propagate the final result to every GPU. With this method, the message size per send/receive in phase 2 is $KM/N$ bytes.

We evaluate the performance of ring-based and hierarchical allreduce algorithms on Cluster-P using FP32 tensors as inputs. NCCL is a library of multi-GPU collective communication primitives, and it contains an efficient implementation of ring-based allreduce. Since there is no open-source implementation of hierarchical allreduce, we use NCCL APIs to implement it and name it NCCL-H. Compared to OpenMPI,  NCCL and NCCL-H could significantly improve the performance of allreduce operations. When the input tensor size is greater than 1MB, NCCL outperforms NCCL-H in all test cases. Intra-group reduce and broadcast operations become the performance bottleneck in NCCL-H, since they are not bandwidth optimal. Note that NCCL-H may have better performance than NCCL if the intra-group bandwidth is large enough: for example, GPUs in the same group are connected by NVLink.

We measure the performance of distributed DNN training with ring-based allreduce. In this set of experiments, we use NCCL to perform allreduce in PyTorch and System-I, and measure training throughput on Cluster-P. We also measure the performance of Horovod \cite{sergeev2018horovod} on PyTorch. In addition to ring-based allreduce, Horovod has other network optimization techniques, such as tensor fusion.  Figure \ref{Fig: Ring_Pytorch_Sensenet} shows the results.

\setlength{\minipagewidth}{0.225\textwidth}
\setlength{\figurewidthFour}{\minipagewidth}
\begin{figure}
    \centering
    \begin{minipage}[t]{\minipagewidth}
    \begin{center}
    \includegraphics[width=\figurewidthFour]{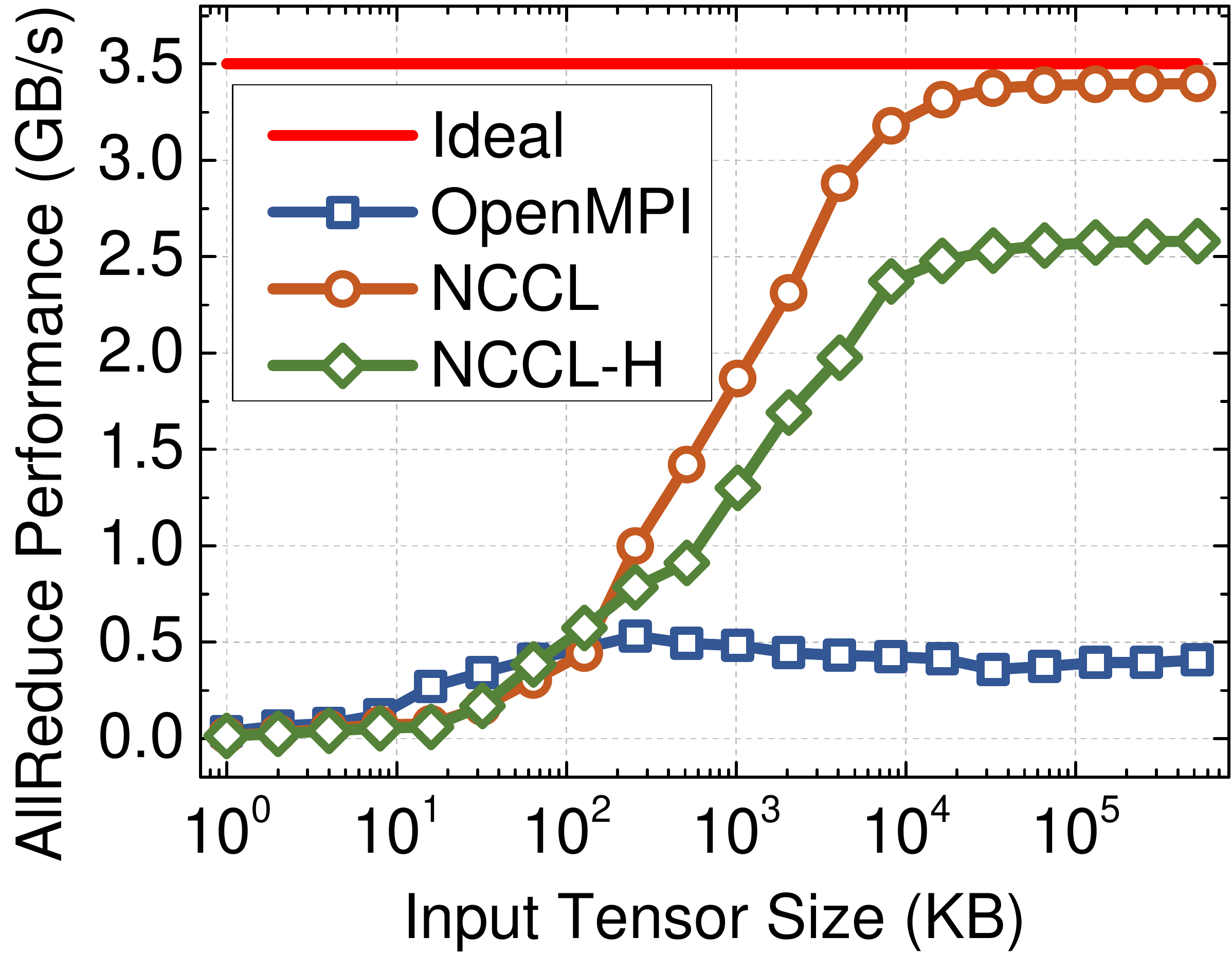}
    \subcaption{(a) 16-GPU}
    \end{center}
    \end{minipage}
    \centering
    \hspace{8pt}
    \begin{minipage}[t]{\minipagewidth}
    \begin{center}
    \includegraphics[width=\figurewidthFour]{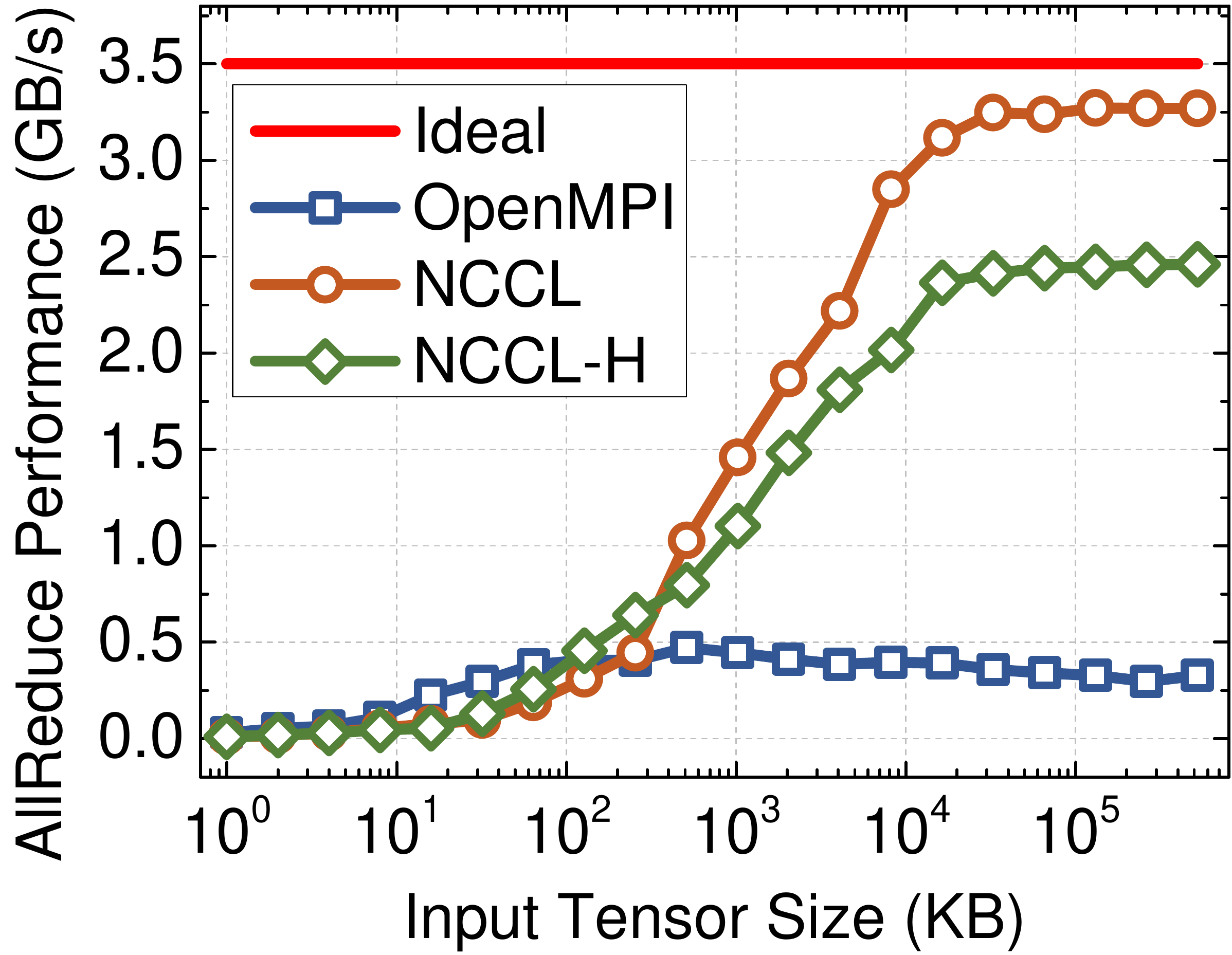}
    \subcaption{(b) 32-GPU}
    \vspace{0pt}
    \end{center}
    \end{minipage}
    \vspace{0pt}
    \centering
    \begin{minipage}[t]{\minipagewidth}
    \begin{center}
    \includegraphics[width=\figurewidthFour]{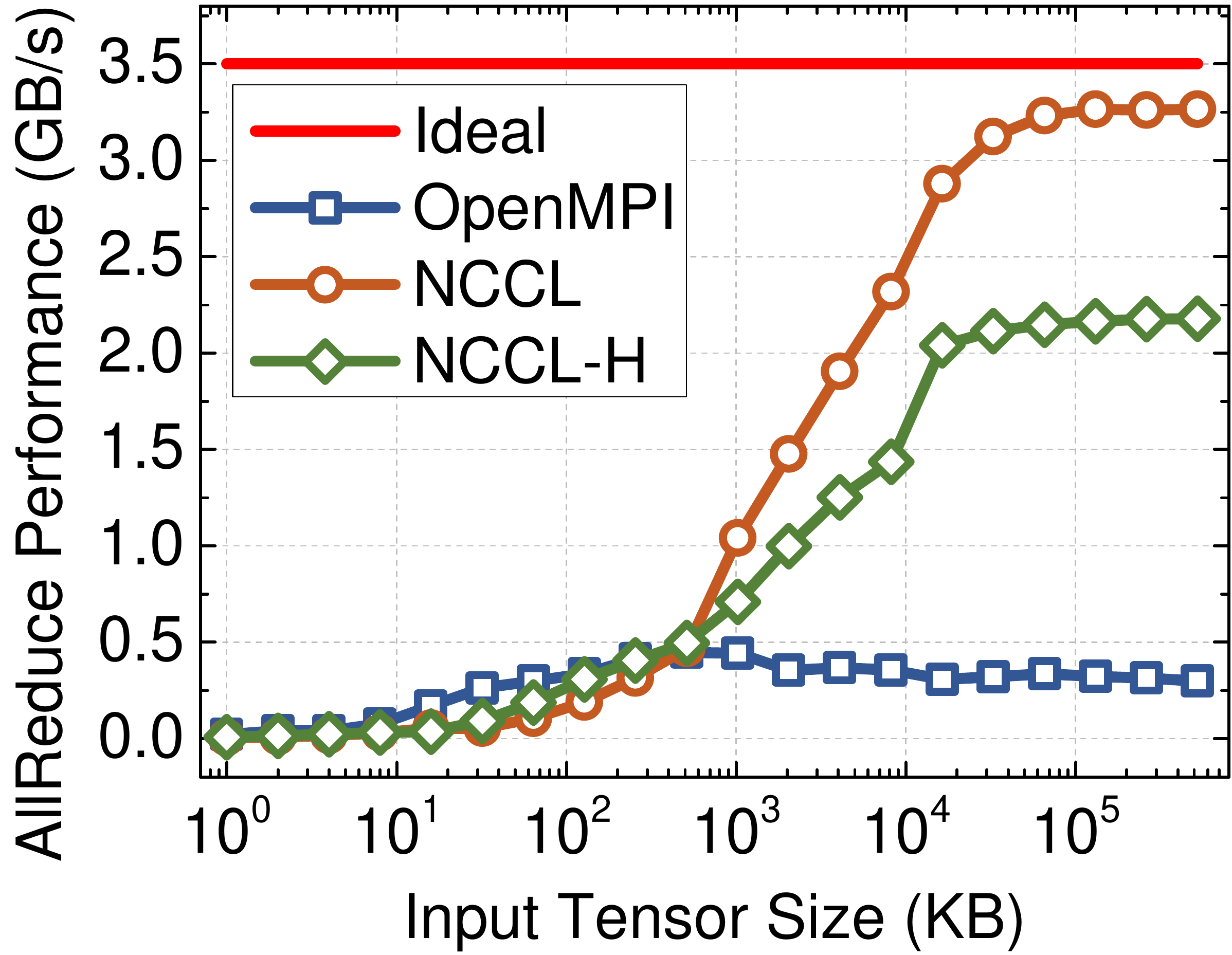}
    \subcaption{(c) 64-GPU}
    \end{center}
    \end{minipage}
    \centering
    \hspace{8pt}
    \begin{minipage}[t]{\minipagewidth}
    \begin{center}
    \includegraphics[width=\figurewidthFour]{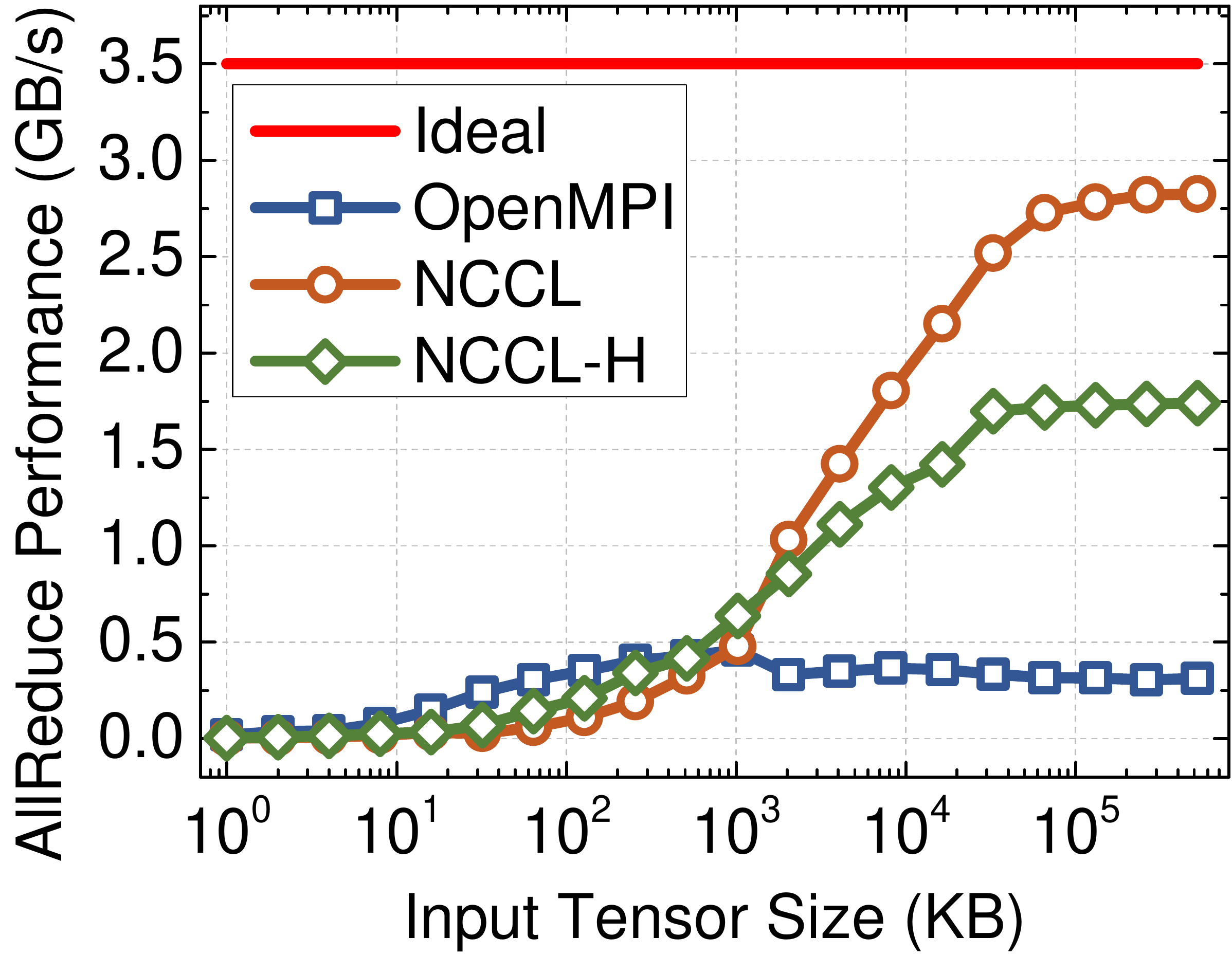}
    \subcaption{(d) 128-GPU}
    \end{center}
    \end{minipage}
    \centering
    \caption{Performance evaluation of FP32 tensor allreduce on Cluster-P with OpenMPI, NCCL and NCCL-H. }
\label{Fig: allreduce_perf}
\end{figure}

\setlength{\minipagewidth}{0.235\textwidth}
\setlength{\figurewidthFour}{\minipagewidth}
\begin{figure}
    \begin{minipage}[t]{\minipagewidth}
    \begin{center}
    \includegraphics[width=\figurewidthFour]{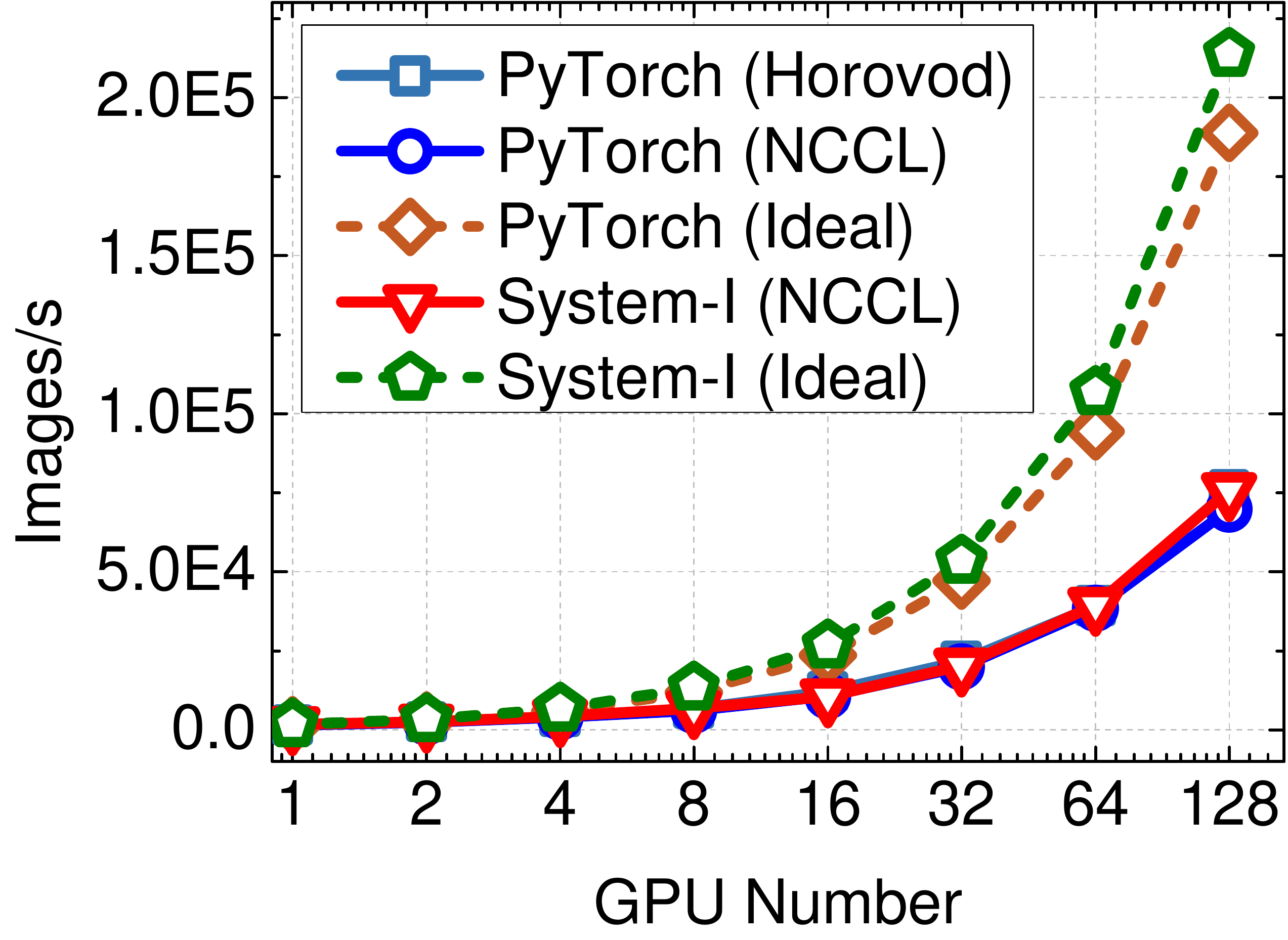}
    \subcaption{(b) AlexNet}
    \end{center}
    \end{minipage}
    \centering
    \begin{minipage}[t]{\minipagewidth}
    \begin{center}
    \includegraphics[width=\figurewidthFour]{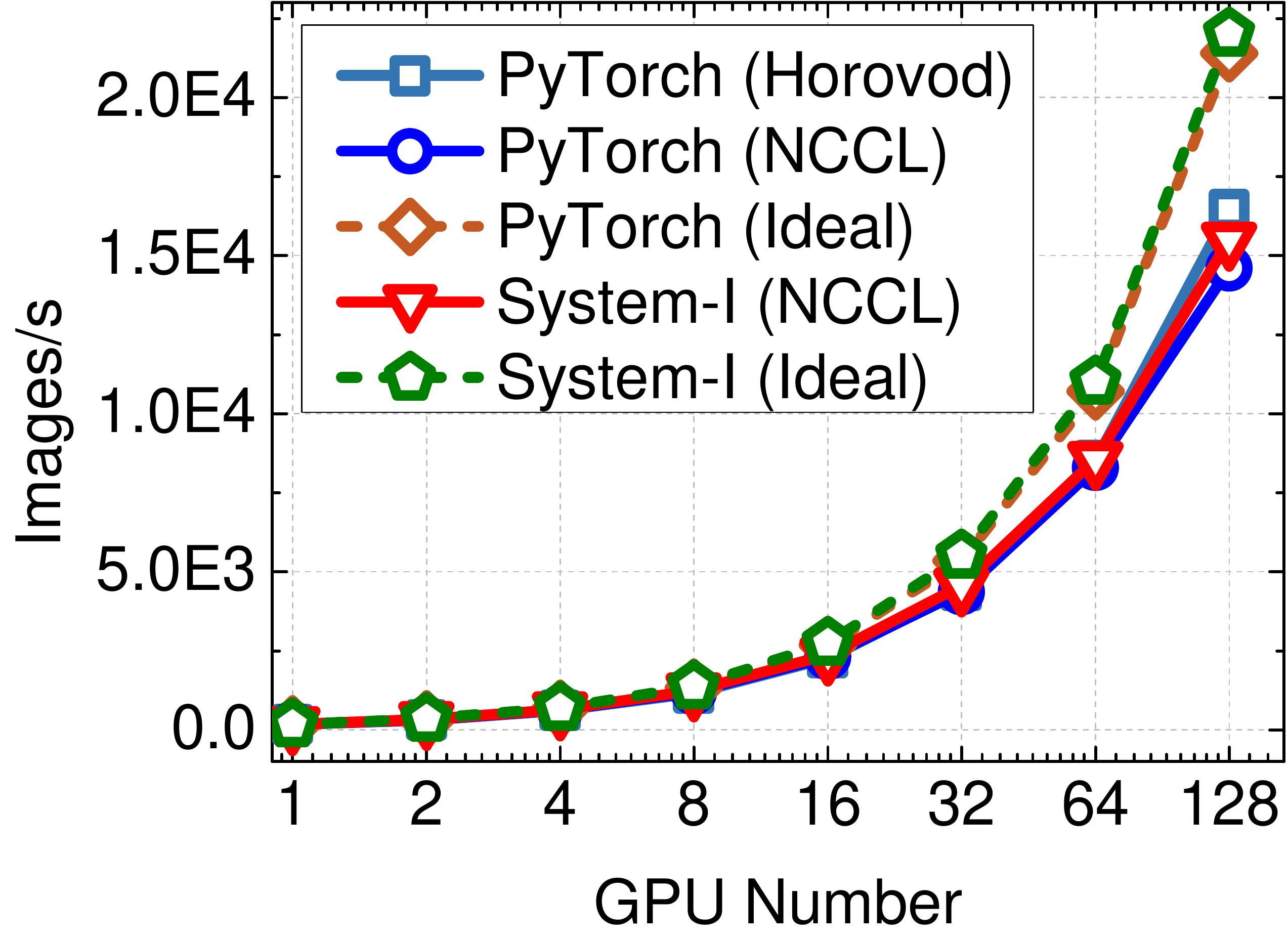}
    \subcaption{(b) ResNet-50}
    \end{center}
    \end{minipage}
    \centering
    \caption{PyTorch and System-I (single-precision) performance evaluation with ring-based allreduce on Cluster-P. The per-GPU batch size is 128 and 64 for AlexNet and ResNet-50.}
\label{Fig: Ring_Pytorch_Sensenet}
\end{figure}

Compared to MPI, NCCL could improve the performance of distributed DNN training to a certain degree. PyTorch (NCCL) can respectively process 69.7K and 14.6K images per second for AlexNet and ResNet-50 with 128 GPUs. The corresponding values for System-I (NCCL) are 75.7K and 15.5K. Compared to MPI, NCCL improves the processing throughput by 3.3x and 1.1x for AlexNet and ResNet-50  on System-I. However, even with NCCL, expensive communication operations still significantly reduce the system performance. On 128 GPUs, PyTorch and System-I with NCCL could only achieve 47.1 and 45.3 speedup ratio for AlexNet training, and achieves 87.3 and 90.1 speedup ratio for ResNet-50 training. While Horovod employs other network optimizations like tensor fusion, it has similar training performance with NCCL. PyTorch (Horovod) could process 75.8K and 16.5K images per second for AlexNet and ResNet-50 training on 128 GPUs. Just using ring-based allreduce could not address the communication challenge of distributed DNN training.

\subsection{Mixed-Precision Training}

Parameters and gradients can be stored in IEEE half-precision (FP16) format instead of single-precision (FP32) format during DNN training. The motivation of using FP16 in the training phase is to lower GPU memory bandwidth pressure, reduce GPU memory storage requirement, and increase arithmetic throughput. Specifically, GPU memory bandwidth pressure and memory storage requirement are reduced since the same number of values could be stored using fewer bits. Half-precision math throughput in GPUs with Tensor Cores is 2x to 8x higher than for single-precision. To avoid accuracy decrease, mixed-precision training  \cite{micikevicius2017mixed} is proposed by using FP16 parameters and gradients in forward/backward computation phases, and using FP32 values in model update phase.

\setlength{\minipagewidth}{0.475\textwidth}
\setlength{\figurewidthFour}{\minipagewidth}
\begin{figure}
    \centering
    \includegraphics[width=\figurewidthFour]{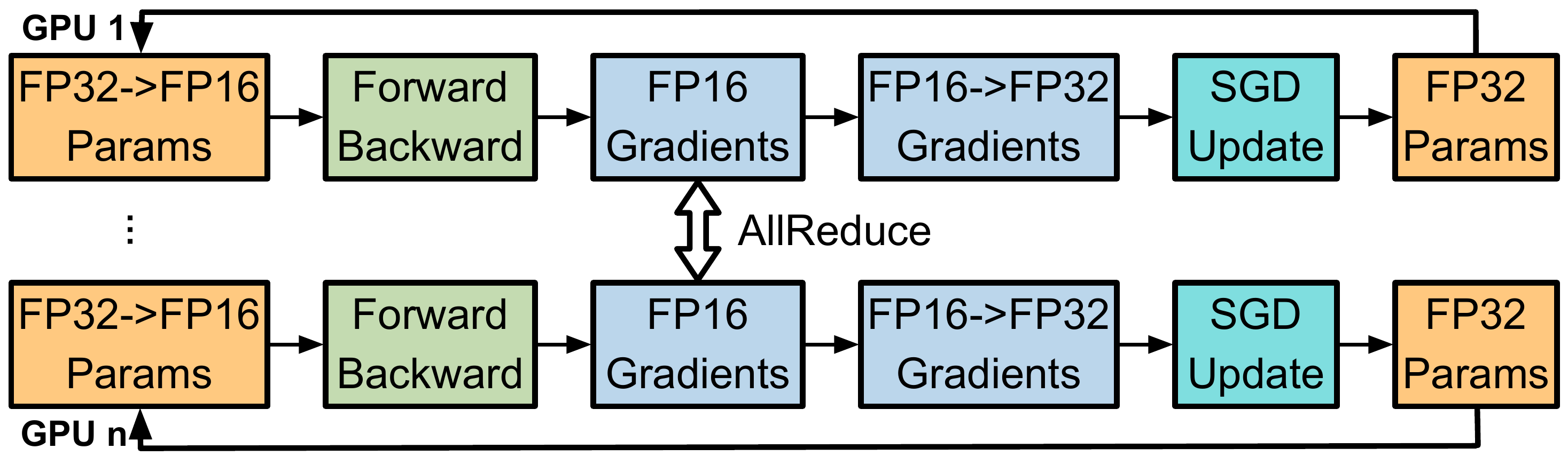}
    \caption{Mixed-precision training uses half-precision gradient tensors as the input of allreduce operations.}
\label{Fig: mixedtrain}
\end{figure}
 \setlength{\minipagewidth}{0.235\textwidth}
\setlength{\figurewidthFour}{\minipagewidth}
\begin{figure}
    \centering
    \begin{minipage}[t]{\minipagewidth}
    \begin{center}
    \includegraphics[width=\figurewidthFour]{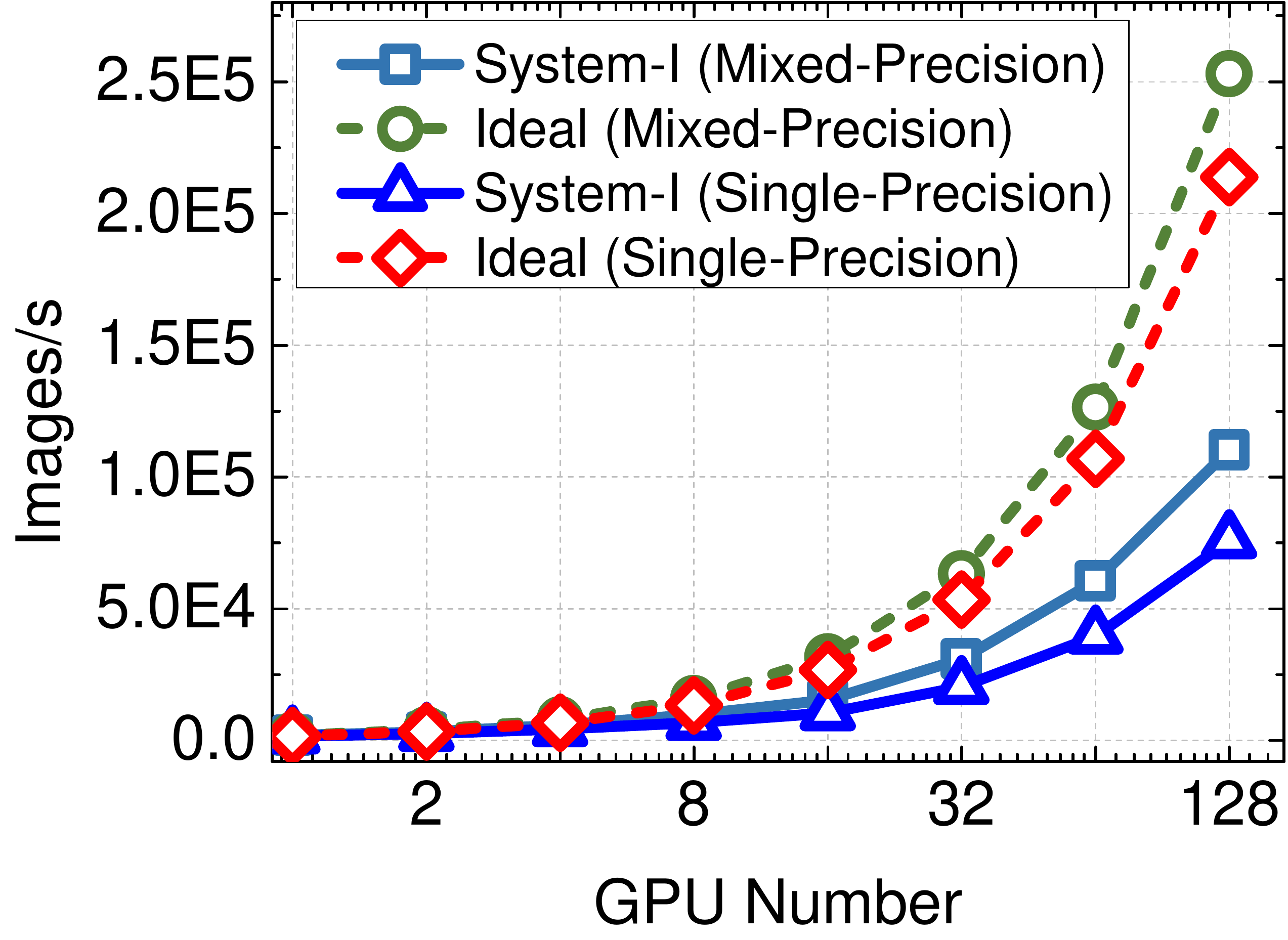}
    \subcaption{(a) AlexNet on Cluster-P}
    \end{center}
    \end{minipage}
    \centering
    \begin{minipage}[t]{\minipagewidth}
    \begin{center}
    \includegraphics[width=\figurewidthFour]{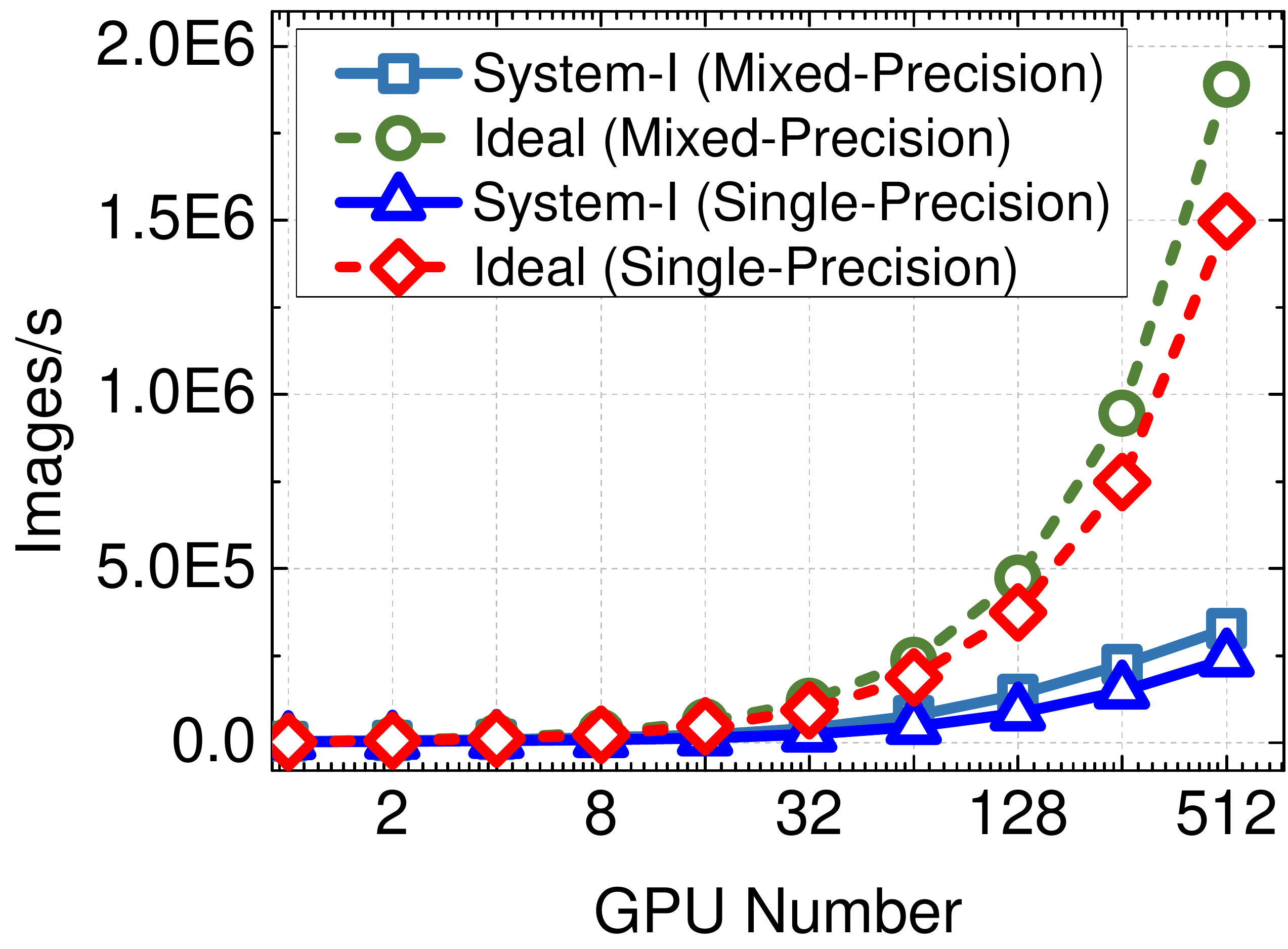}
    \subcaption{(b) AlexNet on Cluster-V}
    \vspace{0pt}
    \end{center}
    \end{minipage}
    \vspace{0pt}
    \centering
    \begin{minipage}[t]{\minipagewidth}
    \begin{center}
    \includegraphics[width=\figurewidthFour]{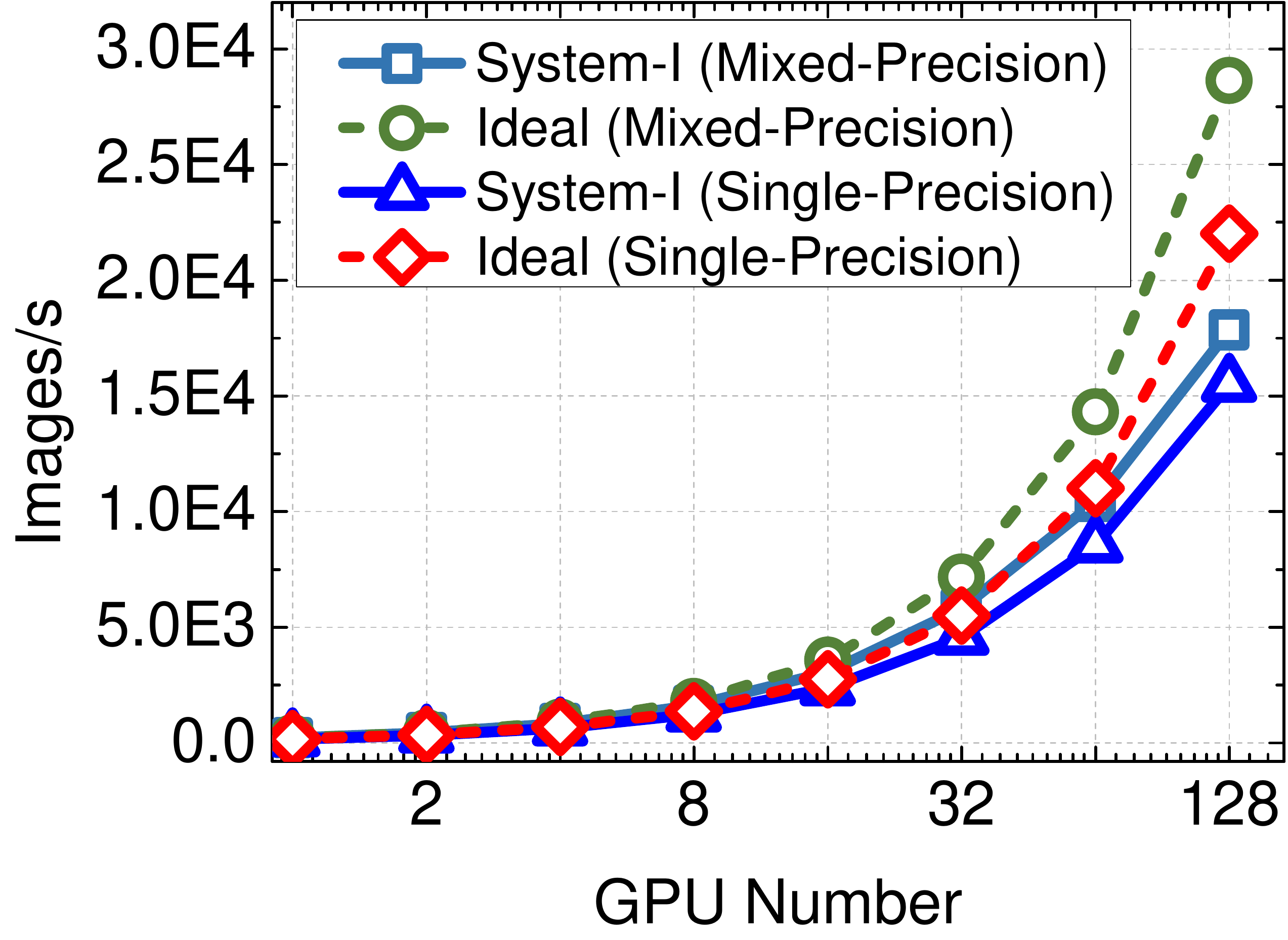}
    \subcaption{(c) ResNet-50 on Cluster-P}
    \end{center}
    \end{minipage}
    \centering
    \begin{minipage}[t]{\minipagewidth}
    \begin{center}
    \includegraphics[width=\figurewidthFour]{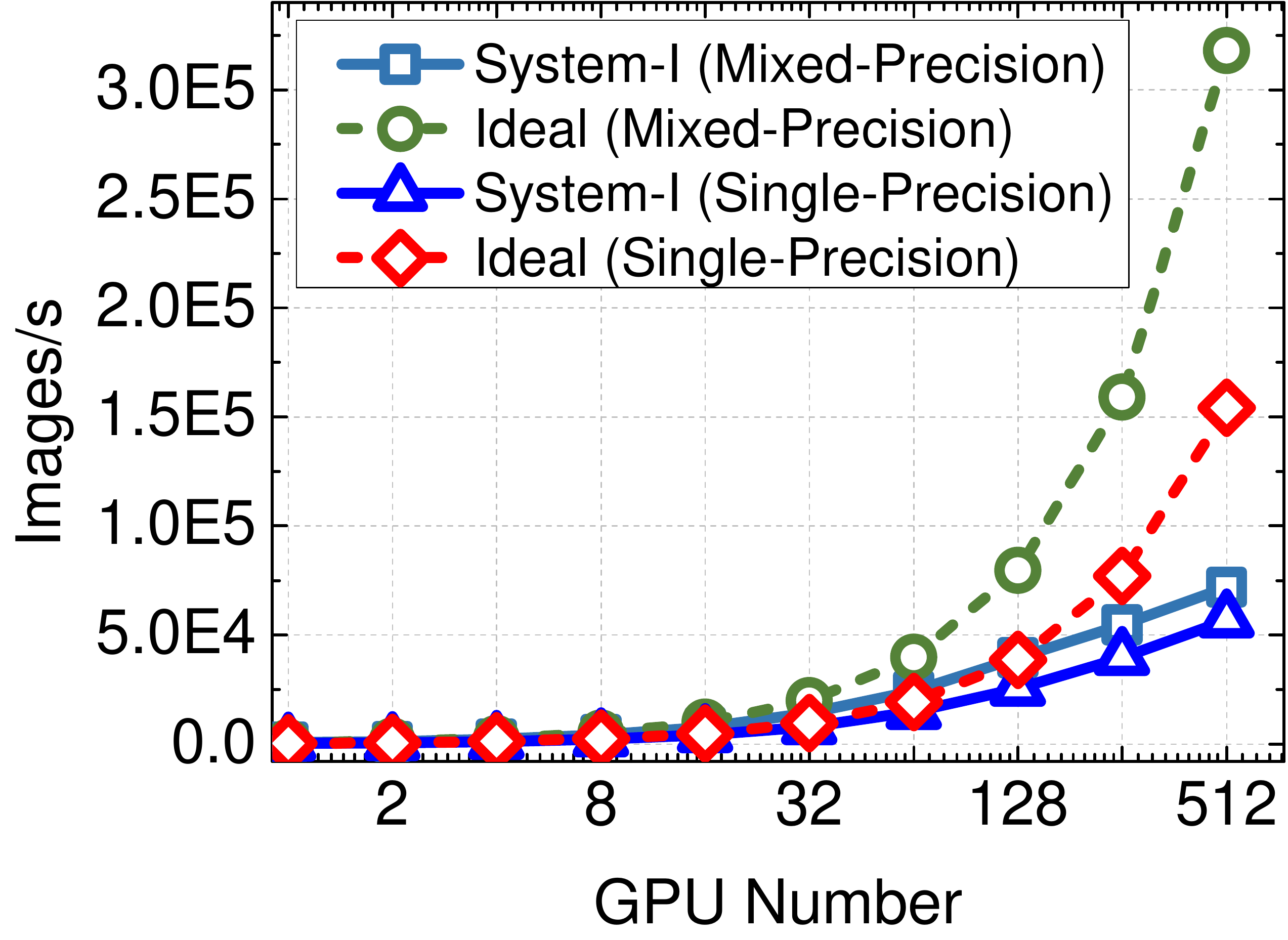}
    \subcaption{(d) ResNet-50 on Cluster-V}
    \end{center}
    \end{minipage}
    \centering
    \caption{System-I performance evaluation on Cluster-P and Cluster-V using single- and mixed-precision training and NCCL. On Cluster-P, the per-GPU batch size is 128 and 64 for  AlexNet and ResNet-50, respectively. On Cluster-V, the per-GPU batch size is 128 for both AlexNet and ResNet-50.}
\label{Fig: mixedp_perf}
\end{figure}

 Micikevicius et al. \cite{micikevicius2017mixed}  runs mixed-precision training on a single GPU. In this paper, we scale mixed-precision training to a distributed setting.  Figure \ref{Fig: mixedtrain} shows that the distributed design of mixed-precision training. We can see that mix-precision also helps to reduce communication overhead, since network transmits fewer bits for the same number of gradients. We implement distributed mixed-precision training strategy in System-I and evaluate its performance on Cluster-P and Cluster-V. When training ResNet-50, Cluster-P only supports 64 per-GPU batch size with FP32 training, and supports 128 per-GPU batch size with mixed-precision training due to reduced memory usage. In this set of experiments, for a fair comparison, the per-GPU batch size is 64 on Cluster-P with mixed-precision training. In addition, we use NCCL as the communication backend. Figure \ref{Fig: mixedp_perf} shows the results.

 Compared to FP32 training, mixed-precision training improves system performance on a single GPU. Specifically, the single-GPU training throughput (images/s) of AlexNet increases to 1.9K from 1.7K on a Pascal GPU, and increases to 3.7K from 2.9K on a Volta GPU. The performance gain of mixed-precision training is not significant for AlexNet, since AlexNet needs to perform 60.9M FP32-to-FP16 and 60.9M FP16-to-FP32 conventions. When training ResNet-50, the single-GPU throughput increases to 223 images/s from 172 images/s on a Pascal GPU, and increases to 621 images/s from 301 images/s on a Volta GPU with the help of Tensor Core.

 Although mix-precision training reduces half of network traffic, the communication overhead is still significant for both AlexNet and ResNet-50 training, as shown in Figure \ref{Fig: mixedp_perf}. When training AlexNet and ResNet-50 with 128 Pascal GPUs, System-I (mixed-precision) only achieves 55.8 and 79.7 speedup ratio, respectively.  When training AlexNet and ResNet-50 with Volta 512 GPUs, System-I (mixed-precision) achieves 88.5 and 115.6 speedup ratio, respectively. We can see that there is still a huge performance gap between System-I with the ideal system with linear speedup ratio.

\subsection{Computation/Communication Overlap}

Distributed DNN training systems can overlap allreduce operations of upper layers with computation of lower layers, reducing dedicated communication time \cite{zhang2015poseidon}. As mentioned in Section 2.3, when finishing the  backward computation of layer-$i$, this layer's gradient tensors are immediately generated, and would not be changed by backward computation of layer-$j$, where $j < i$. Thus,  we can enforce each layer $i$ to start its communication once its gradient tensors are generated. This layer's allreduce operation could be overlapped with the backward computation of layer-$j$, where $j < i$, as shown in Figure \ref{Fig: overlaptrain}. We call this strategy layer-based computation/communication overlap.

 \setlength{\minipagewidth}{0.475\textwidth}
\setlength{\figurewidthFour}{\minipagewidth}
\begin{figure}
    \centering
    \includegraphics[width=\figurewidthFour]{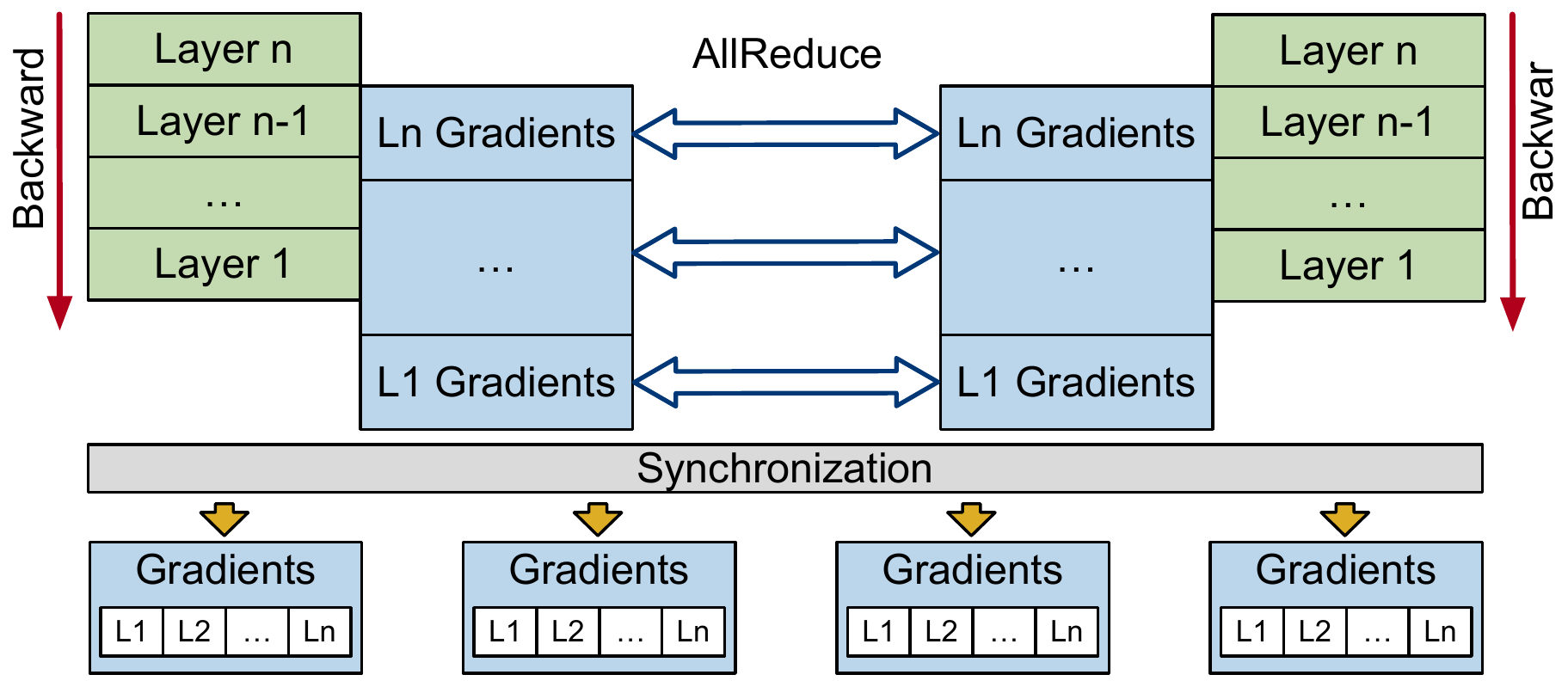}
    \caption{Layer-based computation/communication overlap. In this method, upper layer's communication operations could be overlapped with lower layers' computation operations, thus reducing dedicated communication time of an iteration.}
\label{Fig: overlaptrain}
\end{figure}

Both AlexNet and ResNet-50 can benefit from layer-based computation/communication overlap.  Figure \ref{Fig: layertime} illustrates the average backward computation time of each layer on Volta GPU with mixed-precision training method. For AlexNet with 27 layers, the top 8 layers generate $96.2\%$ of gradients, and consume $7.1\%$ of backward computation time. For ResNet-50 with 188 layers, the top 20 layers generate $56.3\%$ of gradients, and consume $8.9\%$ of backward computation time. Thus, it is possible to reduce dedicated communication time by overlapping upper layers' communication with lower layer's computation. We implement layer-based computation/communication overlap in System-I, and measure its performance on Cluster-P and Cluster-V. We use mixed-precision training and NCCL in this set of experiments.

\setlength{\minipagewidth}{0.475\textwidth}
\setlength{\figurewidthFour}{\minipagewidth}
\begin{figure}
    \centering
    % \begin{minipage}[t]{\minipagewidth}
    % \begin{center}
    % \includegraphics[width=\figurewidthFour]{figure/alex_layer_time_fp32}
    % \end{center}
    % \end{minipage}
    % \centering
    % \\[5pt]
    \begin{minipage}[t]{\minipagewidth}
    \begin{center}
    \includegraphics[width=\figurewidthFour]{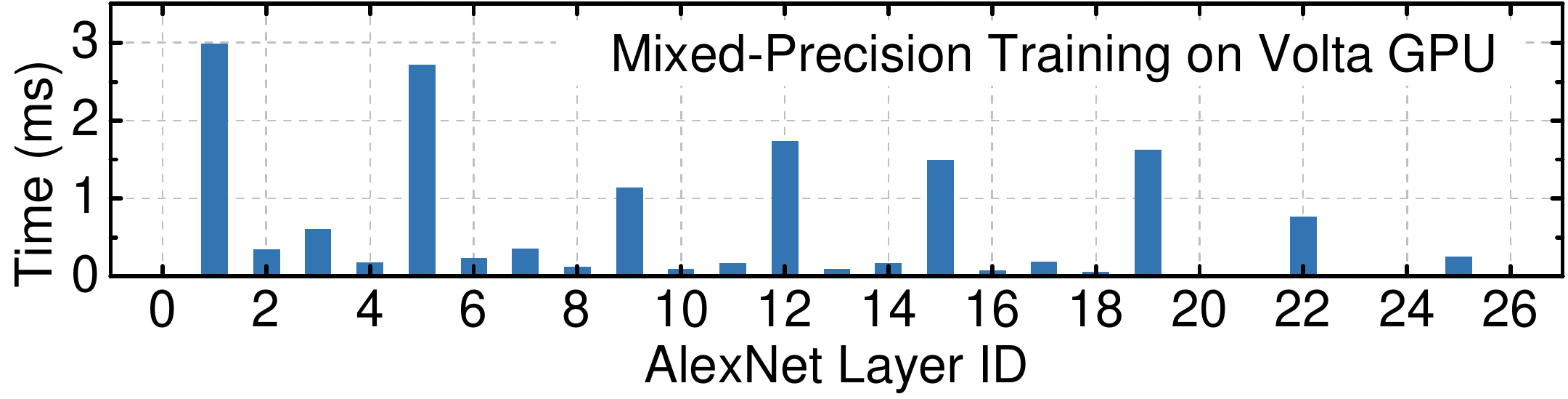}
    \end{center}
    \end{minipage}
    \centering
    % \\[5pt]
    % \begin{minipage}[t]{\minipagewidth}
    % \begin{center}
    % \includegraphics[width=\figurewidthFour]{figure/res_layer_time_fp32}
    % \end{center}
    % \end{minipage}
    % \centering
    \\[5pt]
    \begin{minipage}[t]{\minipagewidth}
    \begin{center}
    \includegraphics[width=\figurewidthFour]{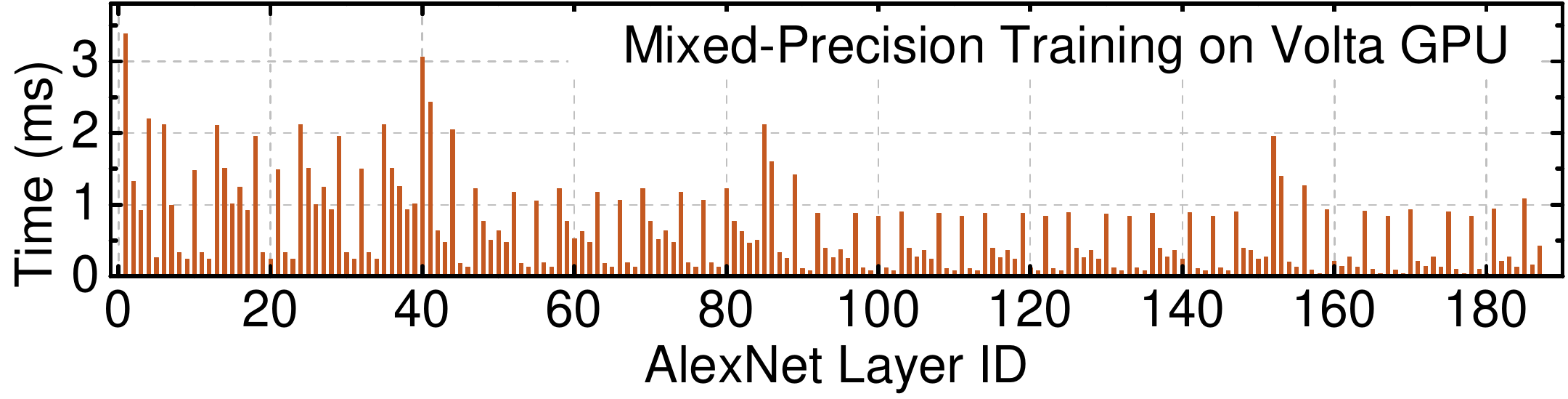}
    \end{center}
    \end{minipage}
    \centering
\caption{Single-GPU, per-layer backward computation time of AlexNet and ResNet-50 on a single Volta GPU.}
\label{Fig: layertime}
\end{figure}

\setlength{\minipagewidth}{0.235\textwidth}
\setlength{\figurewidthFour}{\minipagewidth}
\begin{figure}
    \centering
    \begin{minipage}[t]{\minipagewidth}
    \begin{center}
    \includegraphics[width=\figurewidthFour]{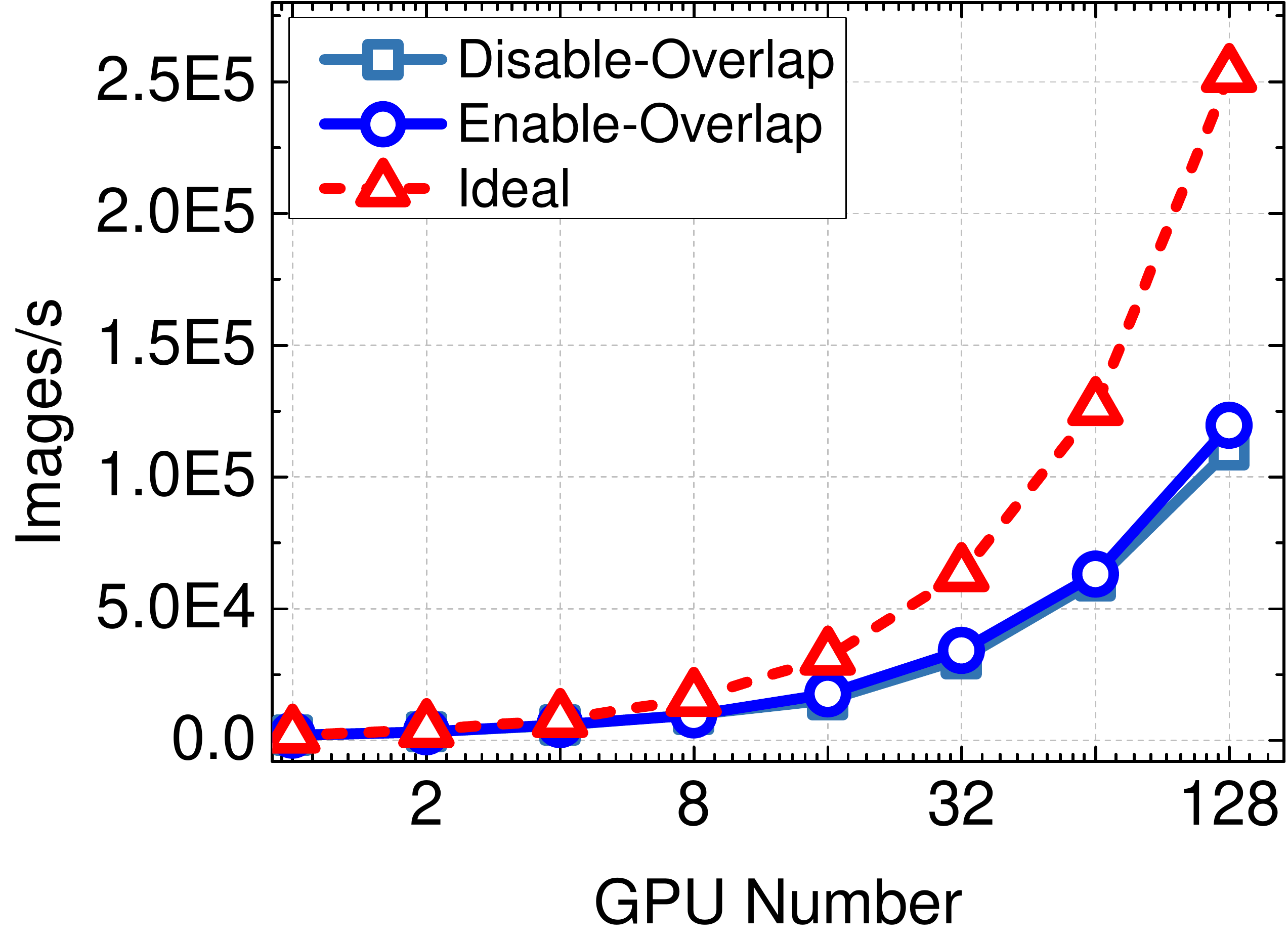}
    \subcaption{(a) AlexNet on Cluster-P}
    \end{center}
    \end{minipage}
    \centering
    \begin{minipage}[t]{\minipagewidth}
    \begin{center}
    \includegraphics[width=\figurewidthFour]{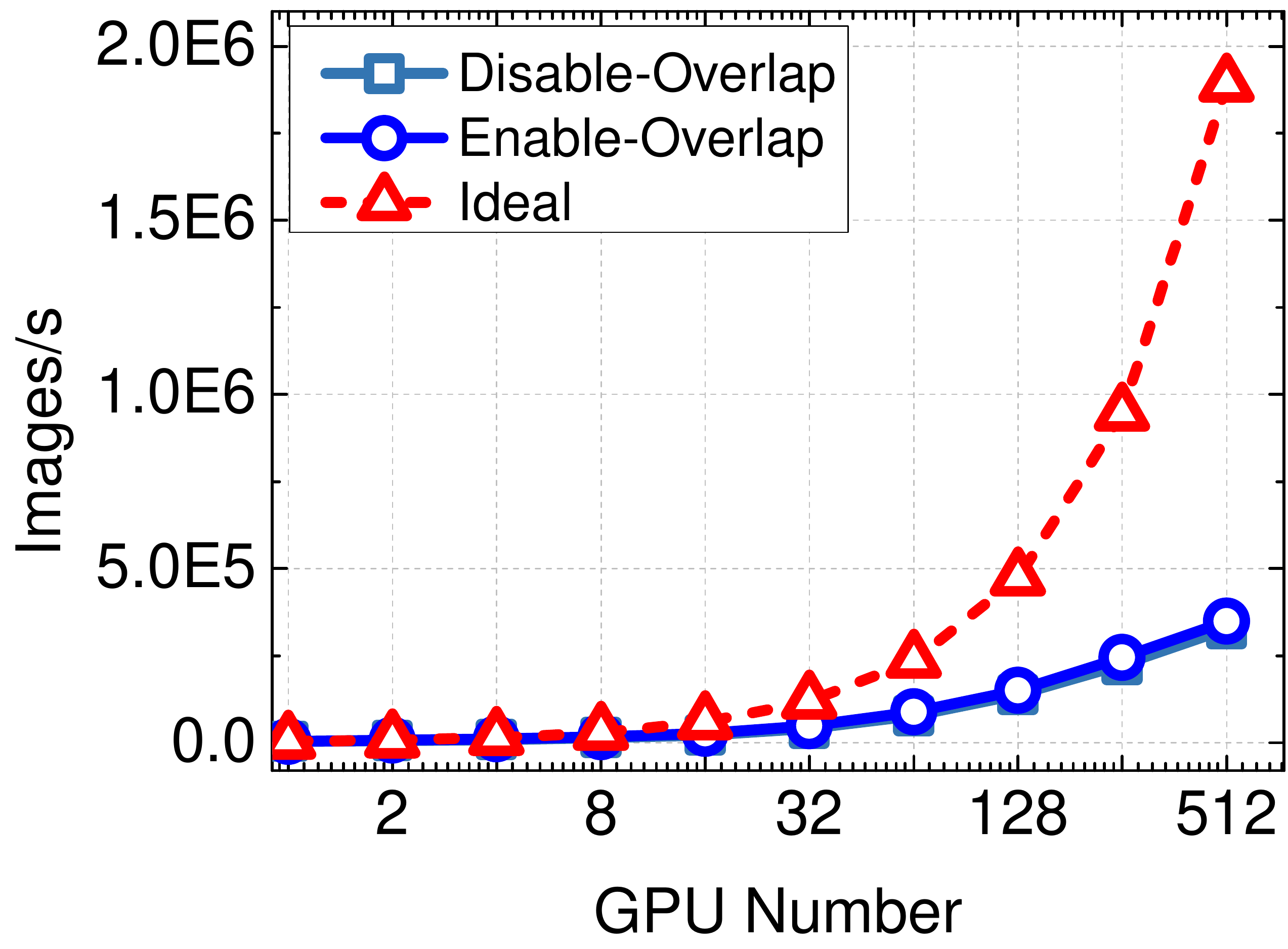}
    \subcaption{(b) AlexNet on Cluster-V}
    \vspace{0pt}
    \end{center}
    \end{minipage}
    \vspace{0pt}
    \centering
    \begin{minipage}[t]{\minipagewidth}
    \begin{center}
    \includegraphics[width=\figurewidthFour]{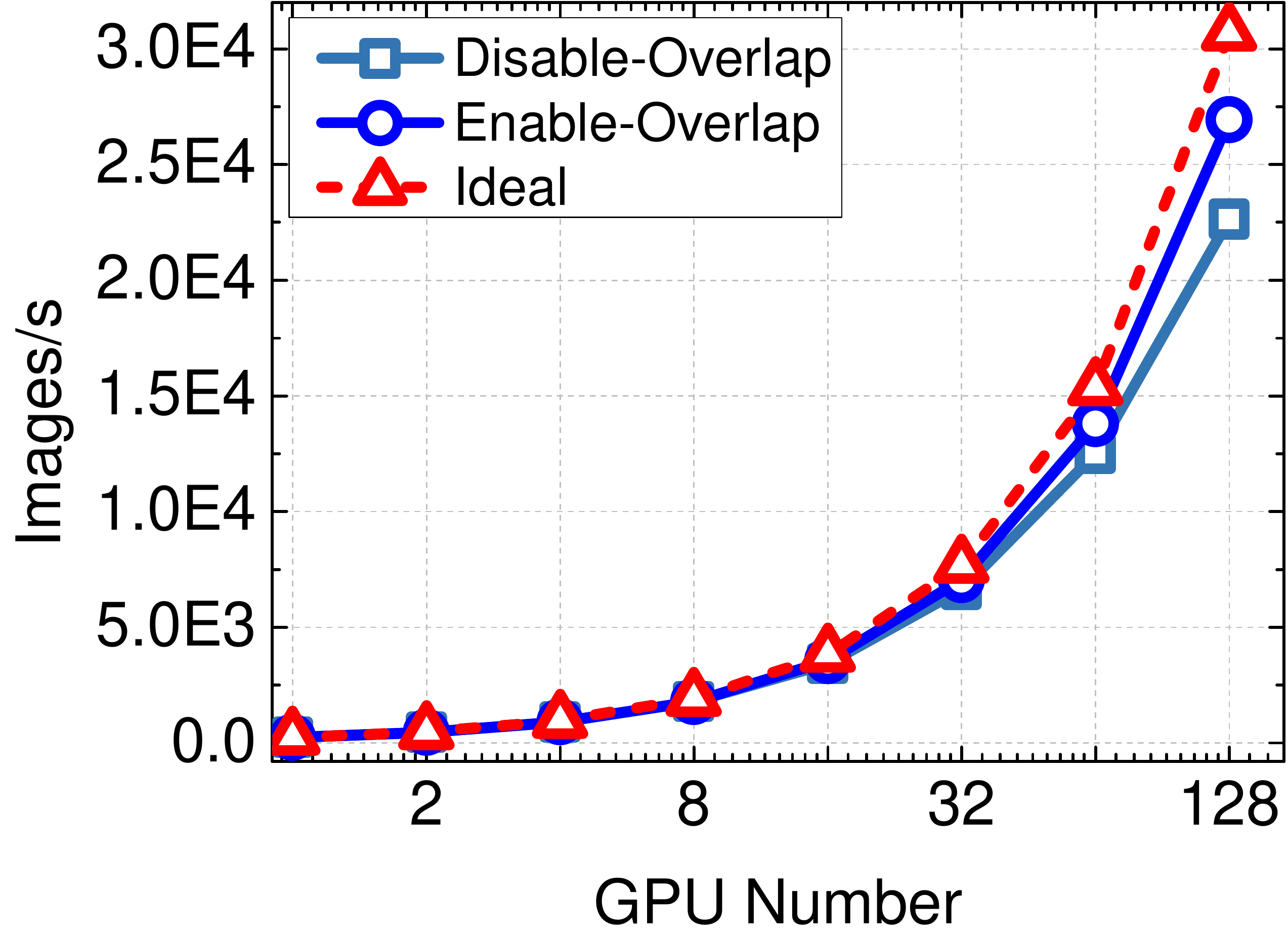}
    \subcaption{(c) ResNet-50 on Cluster-P}
    \end{center}
    \end{minipage}
    \centering
    \begin{minipage}[t]{\minipagewidth}
    \begin{center}
    \includegraphics[width=\figurewidthFour]{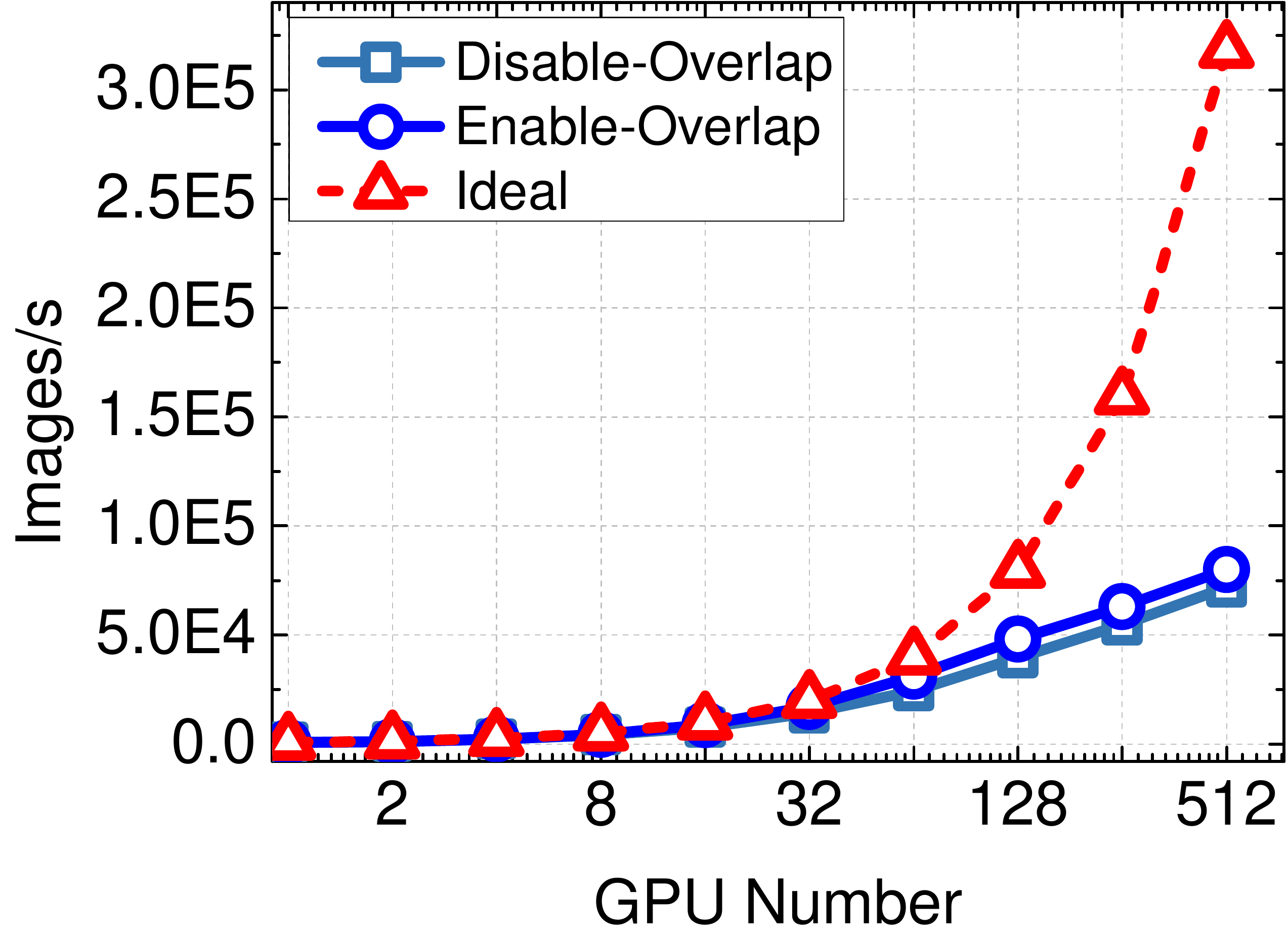}
    \subcaption{(d) ResNet-50 on Cluster-V}
    \end{center}
    \end{minipage}
    \centering
    \caption{System-I (with mixed-precision training and NCCL) performance evaluation on Cluster-P and Cluster-V using layer-based computation/communication overlap. The per-GPU batch size is 128 for AlexNet and ResNet-50.}
\label{Fig: overlap_perf}
\end{figure}

Layer-based computation/communication overlap only improves the performance of distributed DNN training to a certain degree. As shown in Figure \ref{Fig: overlap_perf}, when training AlexNet on 512 Volta GPUs, System-I with overlap approach could improve the training throughput by a factor of $1.068$ (from 326.7K images/s to 349.1K images/s), compared to System-I with just NCCL and mixed-precision training. The corresponding  speedup ratio for ResNet-50 is $1.057$ (from 718.2K images/s to 759.9K images/s).  System-I, which enables NCCL, mixed-precision training and layer-based computation/communication overlap, could achieve 60.5 and 112.3 speedup ratio for AlexNet and ResNet-50 training on 128 Pascal GPUs, and achieve 94.5 and 134.1 speedup ratio for AlexNet and ResNet-50 training on 512 Volta GPUs. Compared to the ideal system, these achieved speedup ratios for AlexNet and ResNet-50 training are still quite low.

\subsection{Conclusion}

Compared to the MPI-based baseline system design, System-I with ring-based allreduce, mixed-precision training, and computation/communication overlap could respectively improve system throughput by 1.49x and 3.82x for training AlexNet and ResNet-50 on 512 Volta GPUs. However, compared to the ideal system with linear speedup ratio, the communication overhead is still significant, since System-I only achieves $18.5\%$ and $26.2\%$ cluster GPU resource utilization for AlexNet and ResNet-50 training on Cluster-V. There are still two problems not addressed: 1) the network utilization is low when performing allreduce on small gradient tensors; 2) the network traffic is huge when training DNNs with a large number of learnable parameters, such as AlexNet.

\section{GradientFlow System Design}

We propose GradientFlow to tackle the high communication cost of distributed DNN training. GradientFlow is a communication backend for System-I, and supports a number of network optimization techniques, including ring-based allreduce, mixed-precision training, and computation/communication overlap. To further reduce network cost, GradientFlow employs lazy allreduce to improve network throughput by fusing multiple allreduce operations into a single one, and employs coarse-graining sparse communication to reduce network traffic by only sending important gradient chucks. In this section, we show the workflow of lazy allreduce and coarse-graining sparse communication, and measure their effectiveness.

\subsection{Lazy AllReduce}

Figure \ref{Fig: DNN-info} shows that AlexNet and ResNet-50 generates a number of small gradient tensors for allreduce in each iteration. NCCL cannot efficiently utilize available network bandwidth to perform allreduce on these small tensors from Figure \ref{Fig: allreduce_perf}. To solve this problem, lazy allreduce tries to fuse multiple allreduce operations into a single one with minimal GPU memory copy overhead. Horovod \cite{sergeev2018horovod} also employs gradient fusion strategy for allreduce. At each iteration, it copies generated gradients into a preallocated fusion buffer, execute the allreduce operation on the fusion buffer, and copy data from the fusion buffer into the output tensor. Due to the additional memory copy operations,  Horovod cannot provider much higher performance than NCCL in our clusters, as shown in section 2.4. Similar performance results of Horovod are reported in \cite{jia2018highly}. Therefore, we need to design a new mechanism to improve network utilization for small tensors.

When a layer with learnable parameters completes its backward computation, it generates one or more gradient tensors. The baseline system allocates a separated GPU memory space for each tensor. With lazy allreduce,  all gradient tensors are placed in a memory pool. As shown in Figure \ref{Fig: lazyallreduce}, a $n$-layer DNN sequentially generates $m$ gradient tensors during the backward phase of an iteration. The elements of all $m$ tensors are placed in a memory pool based on their generated order, from tensor-$m$ to tensor-$1$. The memory pool should  manage $\sum_{i=1}^m\text{sizeof}(\text{tensor-}i)$ FP32 or FP16 gradients in total. When the backward phase computes  tensor-$j$, all its gradients are actually written into the memory pool from offset $\sum_{i=j+1}^m\text{sizeof}(\text{tensor-}i)$ to offset $\sum_{i=j}^m\text{sizeof}(\text{tensor-}i)$.

\setlength{\minipagewidth}{0.475\textwidth}
\setlength{\figurewidthFour}{\minipagewidth}
\begin{figure}
    \centering
    \includegraphics[width=\figurewidthFour]{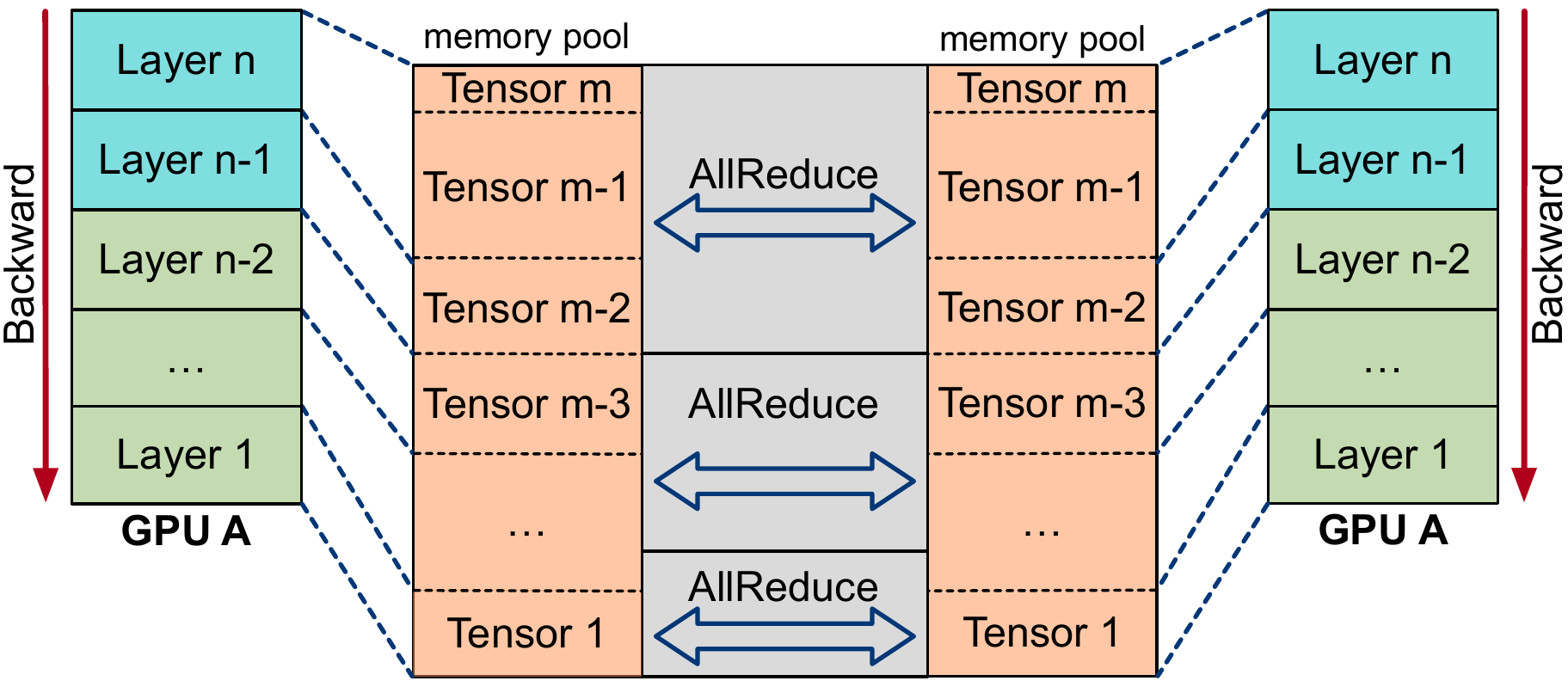}
    \caption{Lazy allreduce would not immediately perform allreduce when a gradient tensor is generated. It tries to fuse multiple allreduce operations into a single one, thus increasing network throughput with large message size. In this example, lazy allreduce overlaps the communication of upper layers with backward computation of lower layers.}
\label{Fig: lazyallreduce}
\end{figure}

\setlength{\minipagewidth}{0.235\textwidth}
\setlength{\figurewidthFour}{\minipagewidth}
\begin{figure}
    \centering
    \begin{minipage}[t]{\minipagewidth}
    \begin{center}
    \includegraphics[width=\figurewidthFour]{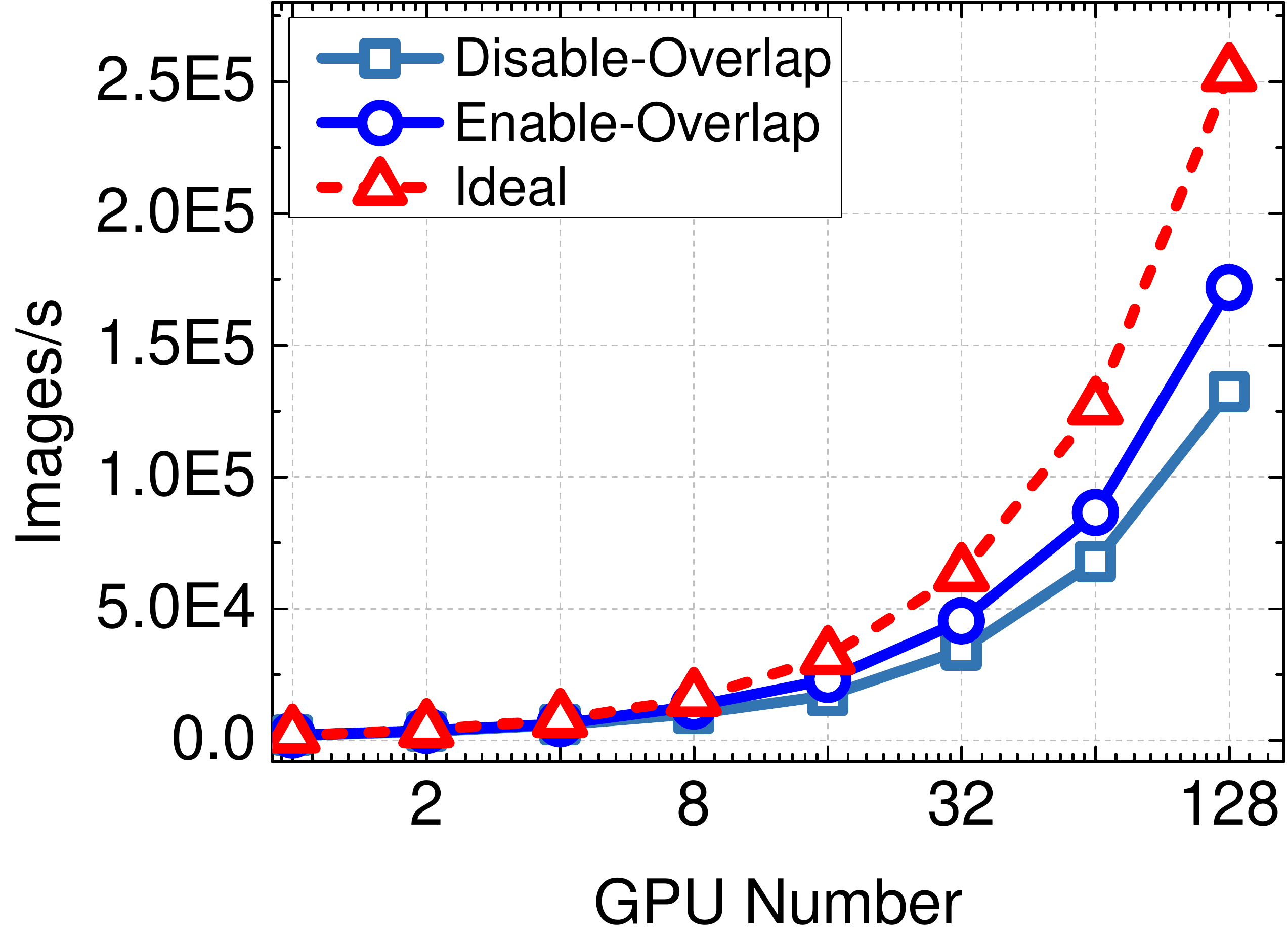}
    \subcaption{(a) AlexNet on Cluster-P}
    \end{center}
    \end{minipage}
    \centering
    \begin{minipage}[t]{\minipagewidth}
    \begin{center}
    \includegraphics[width=\figurewidthFour]{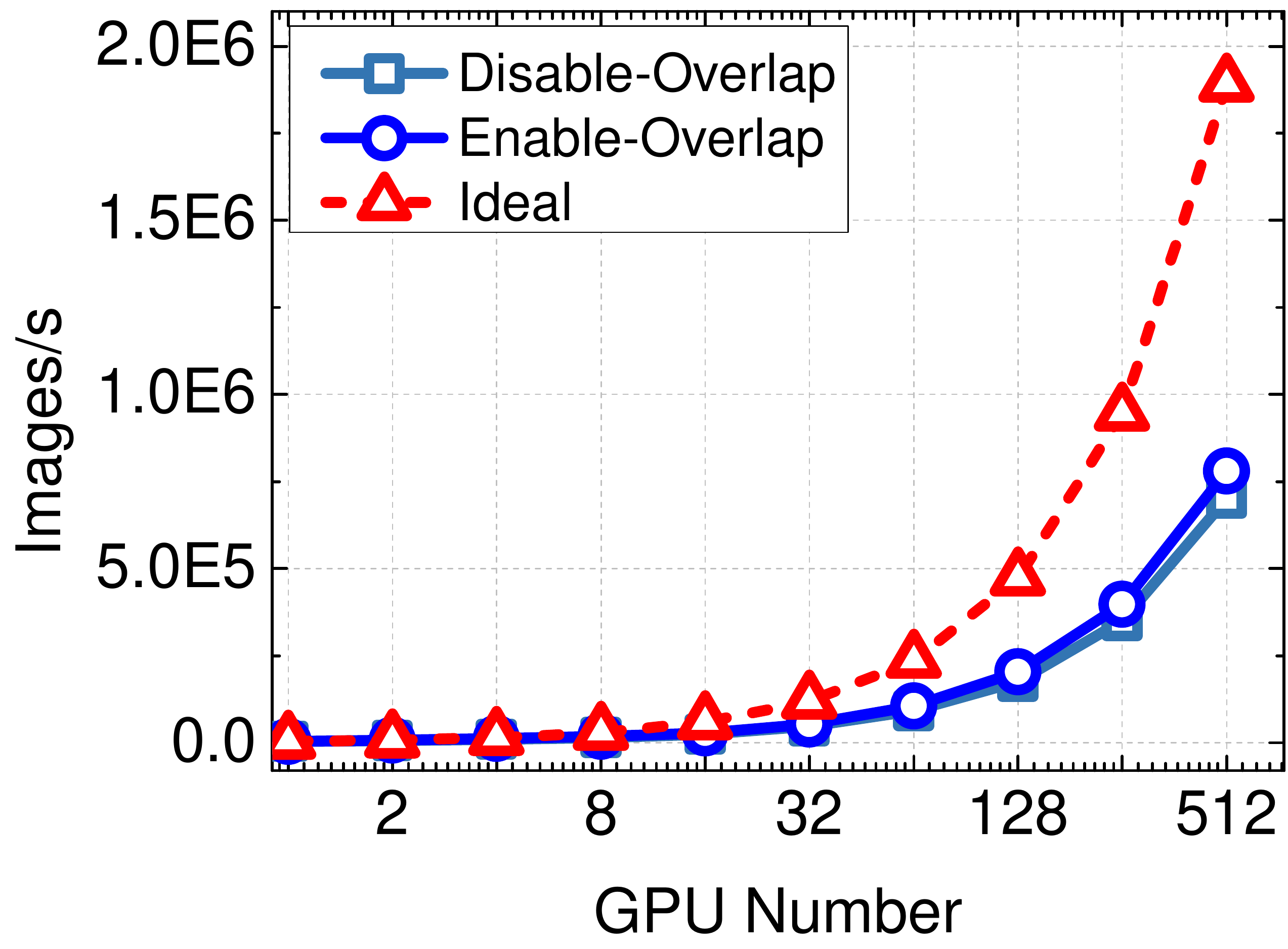}
    \subcaption{(b) AlexNet on Cluster-V}
    \vspace{0pt}
    \end{center}
    \end{minipage}
    \vspace{0pt}
    \centering
    \begin{minipage}[t]{\minipagewidth}
    \begin{center}
    \includegraphics[width=\figurewidthFour]{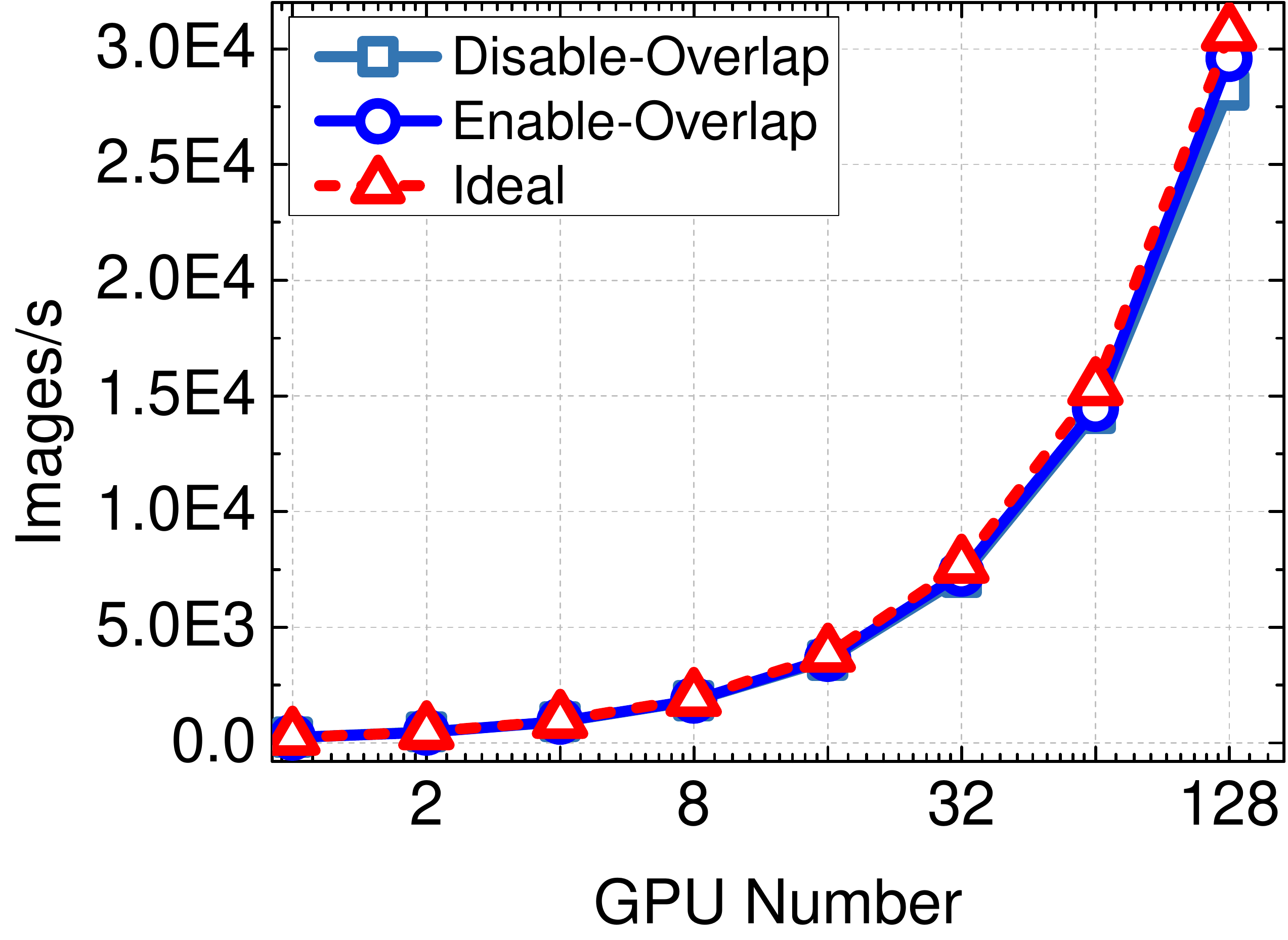}
    \subcaption{(c) ResNet-50 on Cluster-P}
    \end{center}
    \end{minipage}
    \centering
    \begin{minipage}[t]{\minipagewidth}
    \begin{center}
    \includegraphics[width=\figurewidthFour]{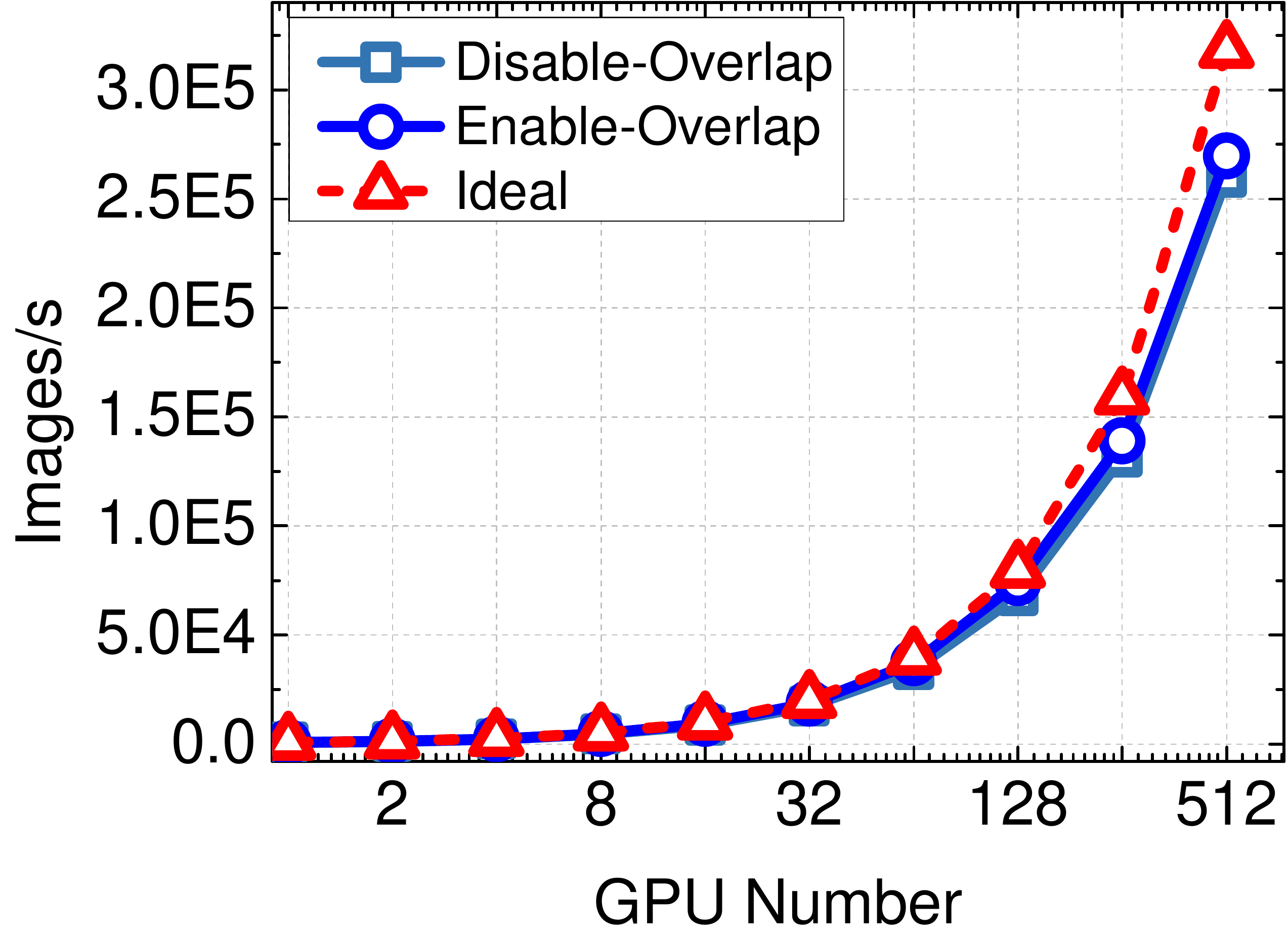}
    \subcaption{(d) ResNet-50 on Cluster-V}
    \end{center}
    \end{minipage}
    \centering
    \caption{System-I performance evaluation with  NCCL,  mixed-precision training and lazy allreduce. The per-GPU batch size is 128. In disable-overlap method, System-I performs a single allreduce operation for all gradient tensors after the backward computation. In enable-overlap method, System-I performs multiple allreduce operations, and could overlap parts of communication with computation.}
\label{Fig: dense_perf}
\end{figure}

Instead of immediately performing allreduce for a generated gradient tensor, lazy allreduce would wait for lower layer's gradient tensors, until the total size of waited tensors is greater than a given threshold $\theta$. Then, we perform a single allreduce operation on all waited gradient tensors. Lazy allreduce thus avoids transmitting small tensors via network and improves network utilization. Since we place all gradient tensors in a memory pool based on their generation order, lazy allreduce could directly perform fused allreduce  on the memory pool, without additional memory copy cost. Compared to Horovod, our design could avoid both network transmission and GPU memory copy for small messages.

We implement lazy allreduce in System-I and evaluate its performance on Cluster-P and Cluster-V along with ring-based allreduce (implemented by NCCL) and mixed-precision training. There are two settings in this set of experiments:  disable-overlap and enable-overlap. In disable-overlap setting, System-I uses a large  communication threshold $\theta$, and performs a single allreduce operation for all gradient tensors after  backward phase. In enable-overlap setting, System-I selects a proper communication threshold $\theta$, and could overlap parts of allreduce operations with backward computation. Figure \ref{Fig: dense_perf} shows the results.

As shown in Figure \ref{Fig: dense_perf}(a)(b), lazy allreduce could help System-I to speed up AlexNet training by 1.2x (from 110.3K images/s to 132.3K images/s) on 128 Pascal GPUs. The corresponding speedup ratio is 2.1 (from 326.7K images/s to 701.1K images/s)  on 512 Volta GPUs. Since AlexNet generates a huge number of network traffic, System-I still has a huge performance gap with the ideal system even with lazy allreduce. In particular, System-I only achieves 211.3 speedup ratio on 512 Volta GPUs for AlexNet training.

Lazy allreduce significantly  improves ResNet-50 training performance by avoiding transmitting small messages. Lazy allreduce (disable-overlap) helps System-I to process 28.2K images per second with 117.7 speedup ratio on 128 Pascal GPUs. The throughput of System-I with lazy allreduce (disable-overlap) is 260.0K images/s with  418.5 speedup ratio on 512 Volta GPUs. If we select a proper communication threshold $\theta$ and enable computation/communication overlap, System-I could process 29.6K images per second with 123.5 speedup ratio on 128 Pascal GPUs, and processes 269.5K images with 434.1 speedup ratio on 512 Volta GPUs.

\subsection{Coarse-Grained Sparse Communication}

Deep gradient compression \cite{lin2017deep} uses fine-grained sparse communication (FCS) to reduce network traffic: only gradients greater than a threshold are selected for transmission.  FCS can reduce up to 99.9\% traffic without losing accuracy. However, FCS runs sparse allreduce with high computation cost and low network utilization, resulting in no significant gains on our clusters. More specifically, each GPU selects different gradients for transmission in FSC, and stores selected values in $k\text{-}v$ format. Given a sparse allreduce operation, each GPU uses its own selected $k\text{-}v$ pairs as input. During allreduce, when a GPU receives a list of $k\text{-}v$ pairs, it accumulates them with its own $k\text{-}v$ pairs, and sends out new $k\text{-}v$ data. Compared to dense algebra operations, these sparse accumulation operations are quite expensive, especially on GPUs. It is also time-consuming to packet a list of $k\text{-}v$ pairs into a buffer for transmission due to the data structure traversal cost and memory copy cost. Our experiments show that an implementation of FCS cannot perform better than NCCL on a physical cluster with 56Gbps network.

\setlength{\minipagewidth}{0.475\textwidth}
\setlength{\figurewidthFour}{\minipagewidth}
\begin{figure}
    \centering
    \includegraphics[width=\figurewidthFour]{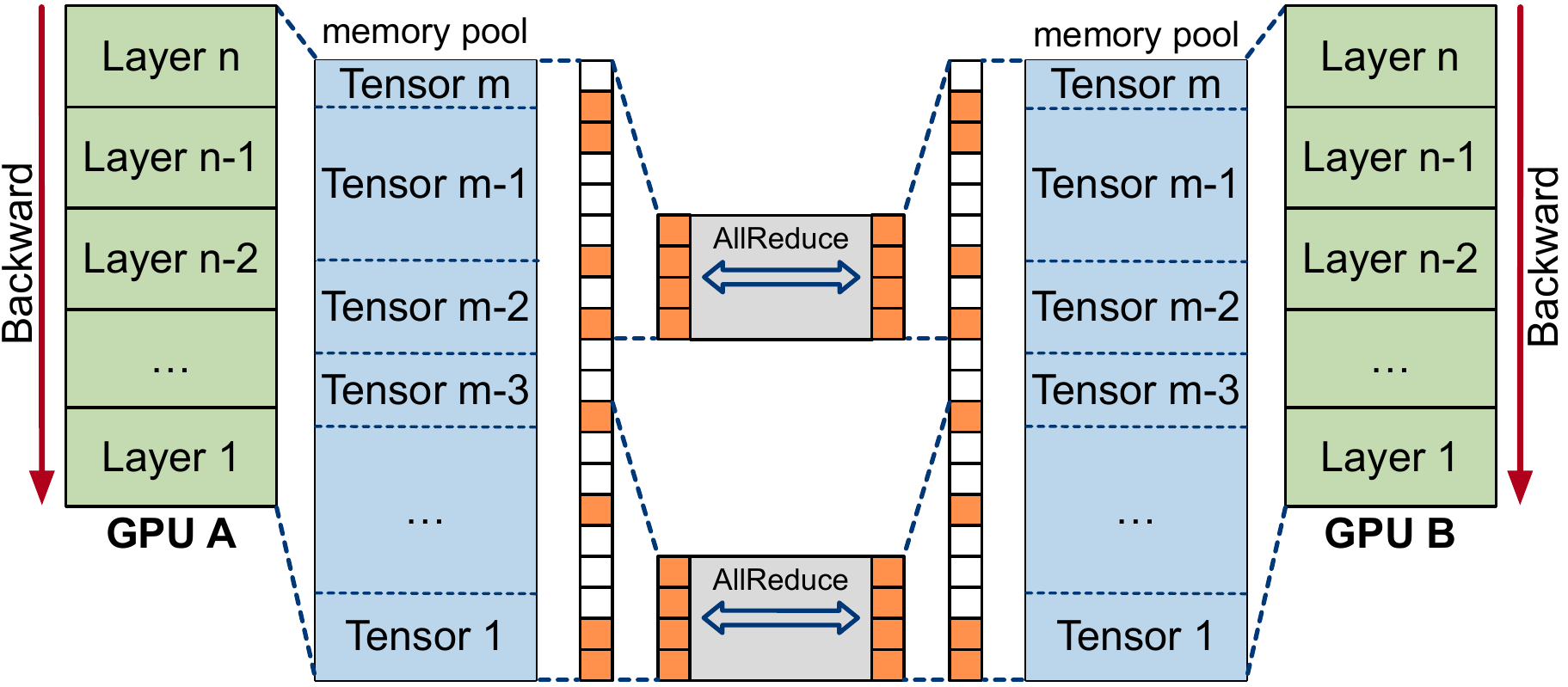}
    \caption{Design of coarse-grained sparse communication (CSC). The gradient memory pool is equally partitioned into a number of small chunks. GPUs could select same important chunks, and copy them into a buffer for allreduce using NCCL. In this example, CSC overlaps the communication of upper layers with backward computation of lower layers.}
\label{Fig: sparsecomm}
\end{figure}

We propose coarse-grained sparse communication (CSC) to reduce network traffic with high bandwidth utilization by selecting important gradient chunks for allreduce. Figure \ref{Fig: sparsecomm} shows the system design of CSC. In CSC, the generated $m$ tensors are also placed in a memory pool with continuance address space based on their generated order. CSC equally partitions the gradient memory pool into chunks, each of which contains a number of gradients. In this work, each chunk contains 32K gradients. In this case, CSC partitions the gradient memory pool of AlexNet and ResNet-50 into 1903 and 797 chunks respectively. A percent (e.g., 10\%) of gradient chunks are selected as important chunks at the end of each iteration. Note that all GPUs should communicate with each other to select same important chunks. CSC also relies on lazy allreduce to avoid transmitting small messages over network. Specifically, when a layer finishes its backward computation, its gradients are written into the corresponding position of the gradient memory pool. If an important chunk is filled, CSC copies its data to a buffer. When the buffer size is greater than a given threshold, CSC starts an allreduce operation using this buffer as input. Since all GPUs select same important chunks, CSC could directly use NCCL to accumulate and broadcast gradients in important chunks with low computation cost and high bandwidth utilization.

 \setlength{\minipagewidth}{0.475\textwidth}
\setlength{\figurewidthFour}{\minipagewidth}
\begin{figure}
    \centering
    \includegraphics[width=\figurewidthFour]{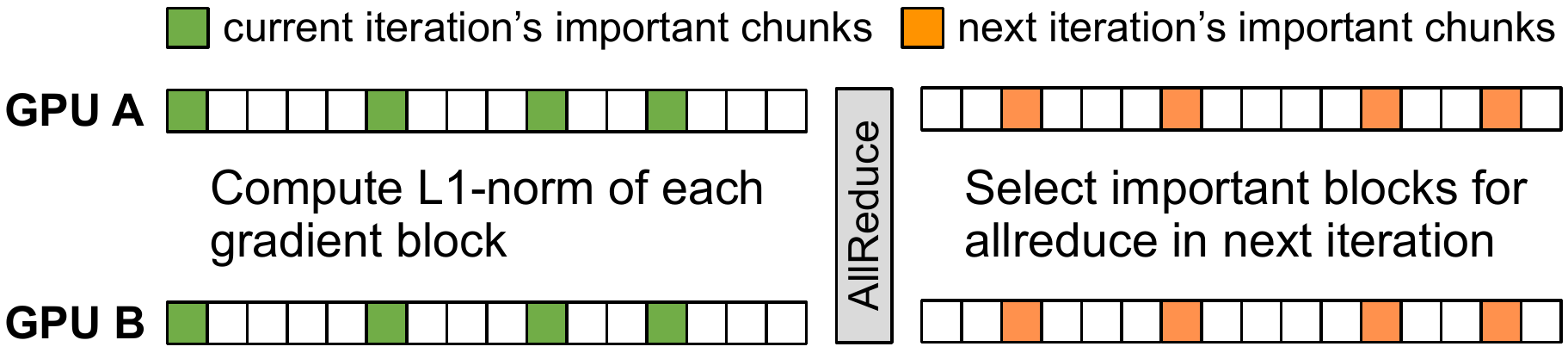}
    \caption{Design of coarse-grained sparse communication.}
\label{Fig: sparseselect}
\end{figure}

\begin{algorithm}
$g_t$: a gradient computed on $t$-th iteration\;
$u_t$: a update computed on $t$-th iteration\;
$w_t$: a weight computed on $t$-th iteration\;
$hg_t$: a historical gradient computed on $t$-th iteration\;
$hu_t$: a historical update computed on $t$-th iteration\;
\textbf{In AllReduce Preprocess Step:}\\
$g_t = g_t + hg_{t-1}$\;
\eIf{this gradient tensor is important}{
$hg_t = 0$\;
}{
$hg_t = momentum * g_t$\;
}
\textbf{In SGD Update Step:}\\
\eIf{this gradient tensor is important}{
$u_t = hu_t = momentum * hu_{t-1} + learningrate * g_t$\;
$w_t = w_t + u_t$;
}{
$hu_t = hu_{t-1}$\;
}
\caption{Momentum SGD Correction}
\label{Alg: sgdcorrect}
\end{algorithm}

Figure \ref{Fig: sparseselect} shows important chunk selection strategy. After the completion of gradient allreduce operations, all GPUs have the same gradients located in important chunks, and keep their own gradients located in other chunks. Then, every GPU computes L1 norm for each chunk. If a chunk is selected as an important chunk in this iteration, an additional computation is needed to divide its L1 norm by the number of GPUs. CSC next performs an allreduce operation for GPUs to exchange and accumulate these L1 norm values.  Finally, GPUs could select a percent of chunks with the largest L1 norms, and mark them as important chunks for the next iteration.

CSC employs momentum SGD correction and warm-up dense training to avoid losing model accuracy.

\textbf{\emph{Momentum SGD Correction. }} Algorithm \ref{Alg: sgdcorrect} the momentum SGD correction algorithm. Given a gradient $g_t$ computed on $t$-th iteration, if it is not located in an important chunk, its value is stored as a historical value: $hg_t = momentum * g_t$. In the following iteration, historical gradients would be accumulated with newly computed gradients before allreduce: $g_t = g_t + hg_{t-1}$. In this way, SCS does not lose any learned information. In momentum SGD update step, if $g_t$ is not in an important chunk, it would not generate an update $u_t$ for updating its corresponding weight $w_t$. Note that the historical update $hu_t$ keeps unchanged with $hu_{t-1}$.

\setlength{\minipagewidth}{0.235\textwidth}
\setlength{\figurewidthFour}{\minipagewidth}
\begin{figure}
    \centering
    \begin{minipage}[t]{\minipagewidth}
    \begin{center}
    \includegraphics[width=\figurewidthFour]{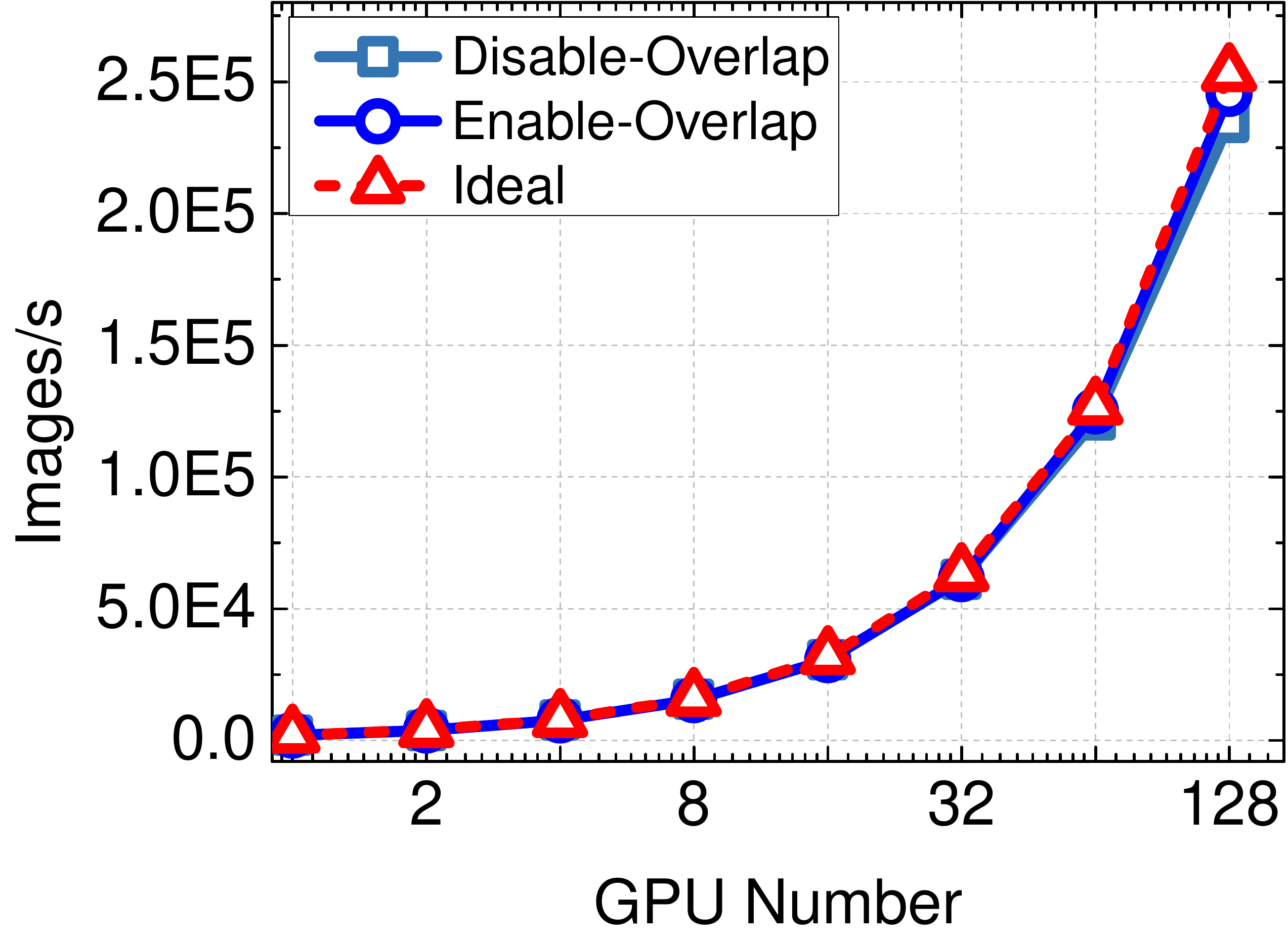}
    \subcaption{(a) AlexNet on Cluster-P}
    \end{center}
    \end{minipage}
    \centering
    \begin{minipage}[t]{\minipagewidth}
    \begin{center}
    \includegraphics[width=\figurewidthFour]{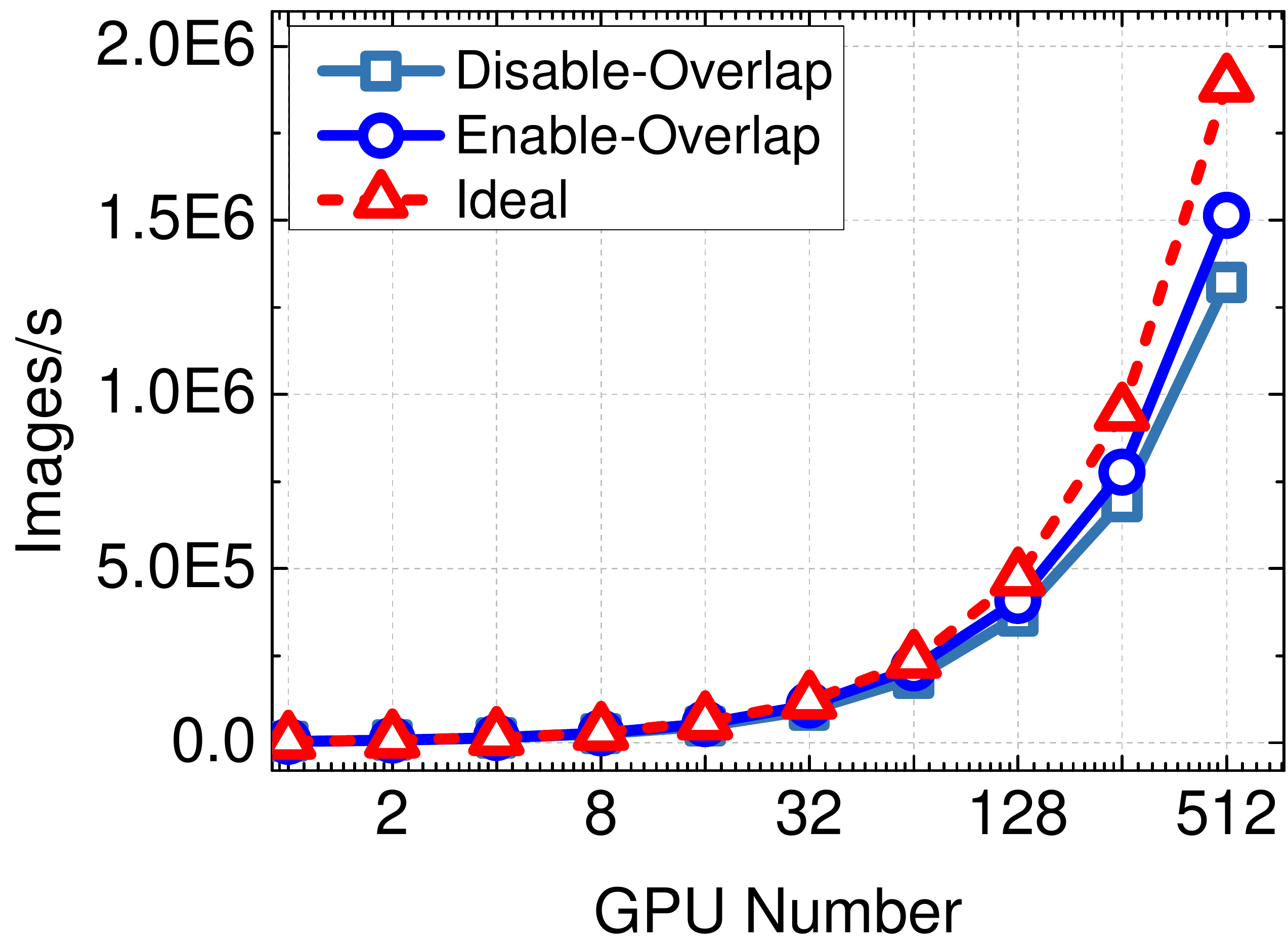}
    \subcaption{(b) AlexNet on Cluster-V}
    \vspace{0pt}
    \end{center}
    \end{minipage}
    \vspace{0pt}
    \centering
    \begin{minipage}[t]{\minipagewidth}
    \begin{center}
    \includegraphics[width=\figurewidthFour]{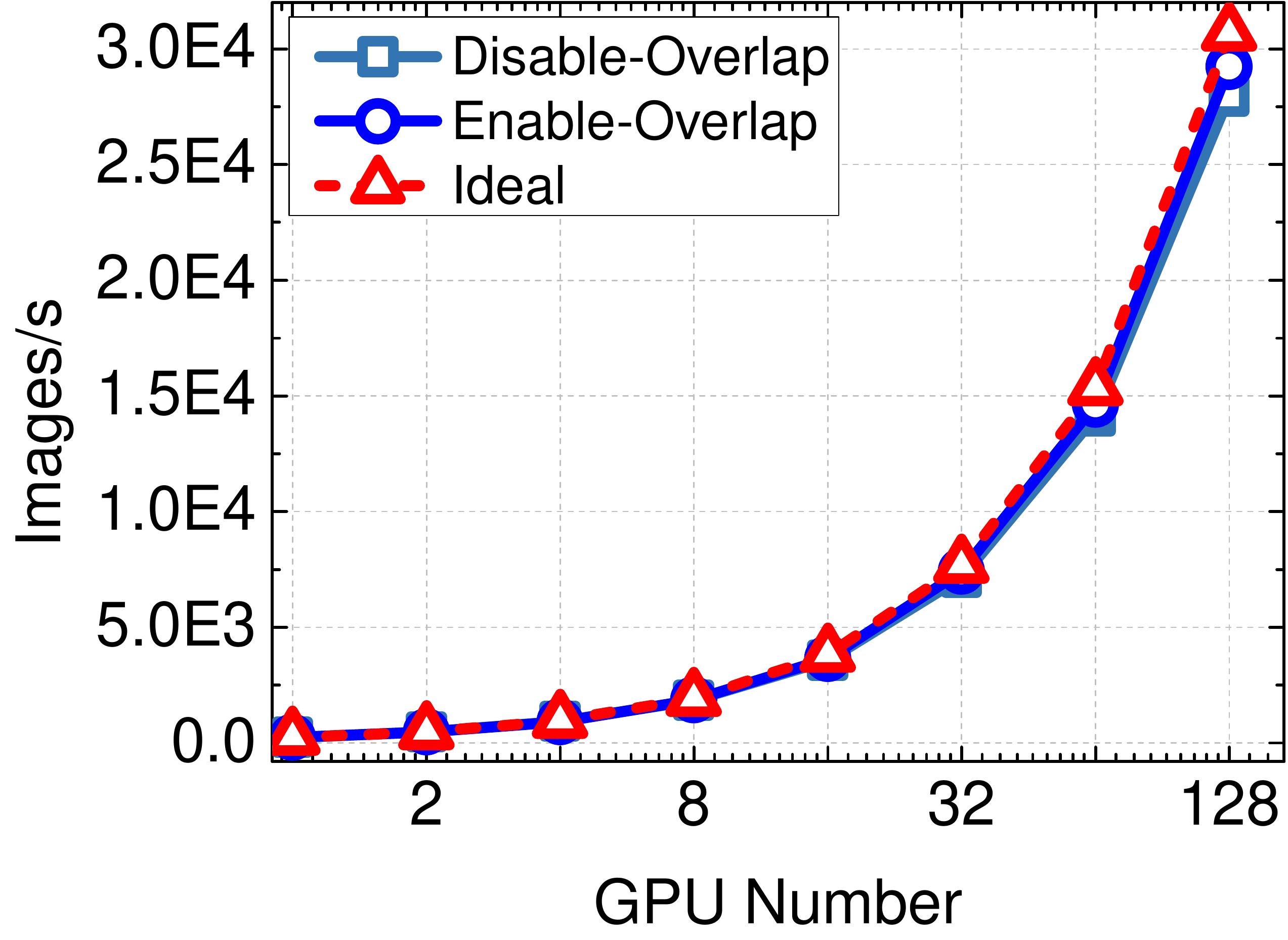}
    \subcaption{(c) ResNet-50 on Cluster-P}
    \end{center}
    \end{minipage}
    \centering
    \begin{minipage}[t]{\minipagewidth}
    \begin{center}
    \includegraphics[width=\figurewidthFour]{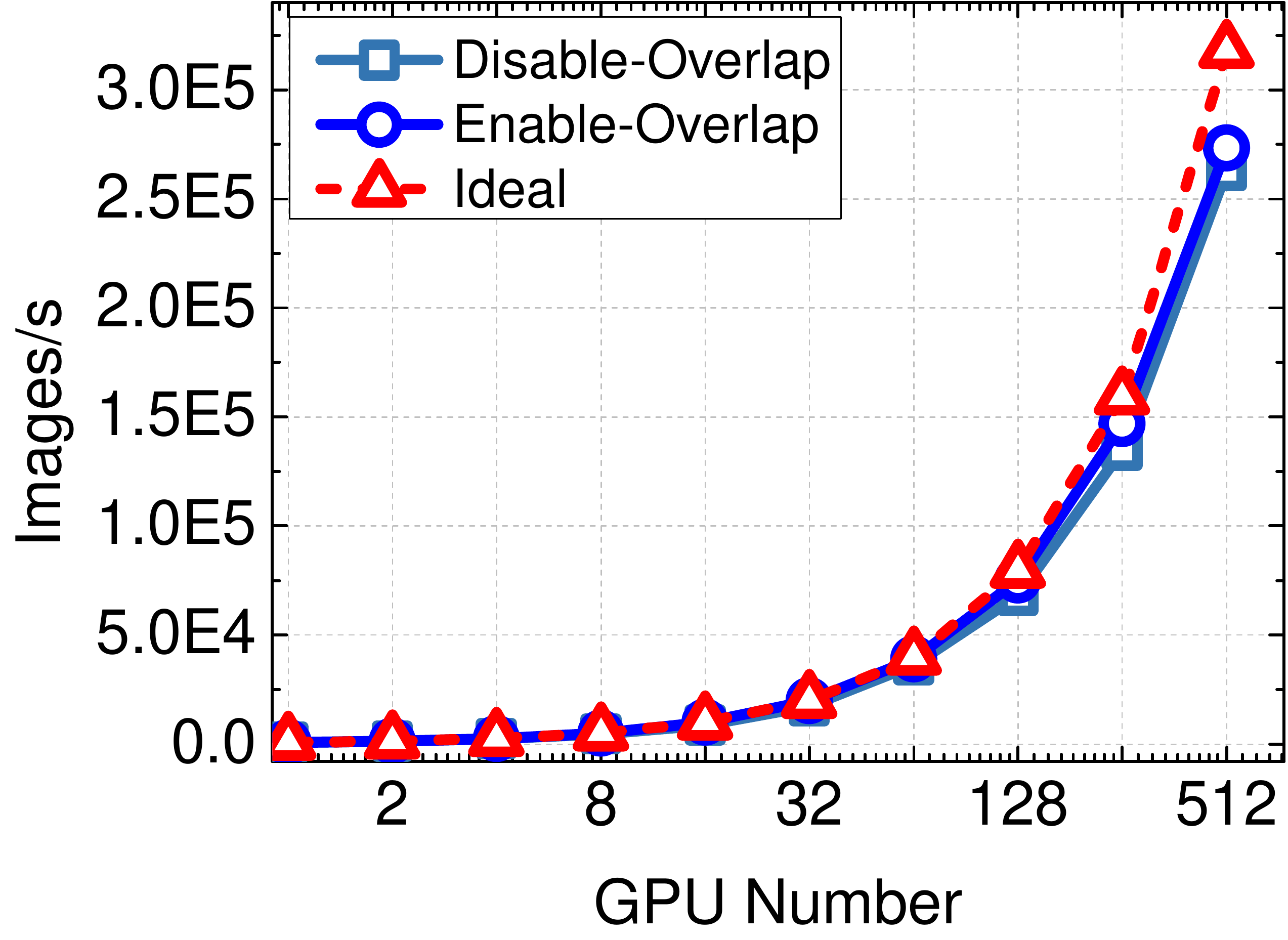}
    \subcaption{(d) ResNet-50 on Cluster-V}
    \end{center}
    \end{minipage}
    \centering
   \caption{System-I performance evaluation with  NCCL,  mixed-precision training and coarse-grained sparse communication. The per-GPU batch size is 128. In disable-overlap method, System-I performs a single allreduce operation for all selected gradient chunks. In enable-overlap method, System-I performs multiple allreduce operations, and could overlap parts of communication with computation.}
\label{Fig: sparse_perf}
\end{figure}

\begin{table*}[]
\centering
\caption{Compare AlexNet training throughput (image/s) and speedup ratio with different network optimizations using System-I on Cluster-P (128 Pascal GPUs) and Cluster-V (512 Volta GPUs). MP: Mixed-Precision Training; LA: Lazy AllReduce; CSC: Coarse-Grained Sparse Communication; Overlap: Computation/Communication Overlap.}
\vspace{-5pt}
\label{Tab: alexperf_commu}
\begin{tabular}{C{5em}C{8em}C{8em}C{8em}C{8em}C{8em}C{8em}}
\toprule
& MPI & NCCL & NCCL+MP &  NCCL+MP +Overlap &  NCCL+MP +LA+Overlap  &  NCCL+MP+LA +CSC+Overlap \\ \midrule
Cluster-P & 22.3K ({1x}) & 75.7K ({3.4x}) &  110.3K ({4.9x}) & 119.5K ({5.4x}) & {172.8K ({7.7x})}  & {245.4K ({11.0x})} \\
Cluster-V & 56.2K ({1x}) & 240.0K ({4.3x}) &  326.7K ({5.8x}) &  349.1K ({6.2x}) & 780.3K ({13.9x}) & 1514.3K ({26.9x})  
\\ \bottomrule
\end{tabular}
\end{table*}

\begin{table*}[]
\centering
\caption{Compare ResNet-50 training throughput (image/s) and speedup ratio with different network optimizations using System-I on Cluster-P (128 Pascal GPUs) and Cluster-V (512 Volta GPUs). MP: Mixed-Precision Training; LA: Lazy AllReduce; CSC: Coarse-Grained Sparse Communication; Overlap: Computation/Communication Overlap.}
\vspace{-5pt}
\label{Tab: resperf_commu}
\begin{tabular}{C{5em}C{8em}C{8em}C{8em}C{8em}C{8em}C{8em}}
\toprule
& MPI & NCCL & NCCL+MP &  NCCL+MP +Overlap &  NCCL+MP +LA+Overlap  &  NCCL+MP+LA +CSC+Overlap \\ \midrule
Cluster-P & 13.9K (1x) & 15.5K (1.1x) & 17.8K (1.3x) &  26.9K (1.9x) & 29.6K (2.1x) &  29.6K (2.1x) \\
Cluster-V & 30.2K (1x)  & 56.8K (1.9x) & 71.8K (2.4x) & 80.0K (2.6x) &  269.5K (8.9x) & 273.2K (9.0x)
\\ \bottomrule
\end{tabular}
\end{table*}

\textbf{\emph{Warm-Up Dense Training. }} In early iterations of training, DNN parameters rapidly change with more aggressive gradients. If we limit the exchange of these aggressive gradients using sparse communication, the DNN training process may have wrong optimization direction. We adopt the warm-up strategy introduced in \cite{lin2017deep}. We set the first several percents of iterations as warm-up iterations. During the warm-up period, we start from dense training with zero sparsity ratio and linearly ramping up the sparsity ratio to the final value.

We implement CSC in System-I and evaluate its performance for Alexnet and ResNet-50 training. Figure \ref{Fig: sparse_perf} shows the average throughput. In this set of experiments, we enable ring-based allreduce and mixed-precision training to maximize the performance.
Due to reduced network traffic and low additional computation cost, CSC significantly improves AlexNet's training  performance, as shown in Figure \ref{Fig: sparse_perf}(a)(b). When enabling computation/communication overlap by selecting a proper communication threshold,  System-I with CSC could process 245.4K images per second, and achieves $124.2$ speedup ratio on 128 Pascal GPUs. The corresponding training throughput on 512 Volta GPUs  is 1514.3K images/s with 410.2 speedup ratio. Compared to System-I with just lazy allreduce, CSC improves the performance of AlexNet training by 1.42x and 1.94x on 128 Pascal GPUs and 512 Volta GPUs, respectively. Since the communication bottleneck of ResNet-50 training is not network traffic, CSC does not improve its training throughout too much. Noted that System-I could already achieve high speedup ratio for ResNet-50 training  with lazy-allreduce.

\section{Overall Performance Evaluation}

In this section, we show the overall performance evaluation results of System-I with GradientFlow to train AlexNet and RestNet-50 with 128 Pascal GPUs and 512 Volta GPUs. We first show the effectiveness of employed network optimization techniques and their combinations. Then we show  System-I's overall training time, and compare it with other approaches.

\subsection{Effectiveness of Network Optimizations}

Table \ref{Tab: alexperf_commu} and Table \ref{Tab: resperf_commu} show the effectiveness of integrated network optimization techniques and their combinations.  NCCL, mixed-precision training and communication/computation overlap could improve training throughput  to a certain degree, but not significantly. Compared to MPI-based baseline system, GradientFlow with NCCL, mixed-precision training and communication/computation overlap would improve the throughput of System-I by 6.2x and 2.6x for AlexNet and ResNet-50 training on 512 Volta GPUs. If GradientFlow enables lazy allreduce, the corresponding speedup ratio increases to 13.9 and 8.9, respectively. Due to reduced network traffic, System-I with coarse-grained sparse communication speeds up AlexNet training by 26.9x on Cluster-V. Since the communication bottleneck of ResNet-50 training is not network traffic, System-I with coarse-grained sparse communication has similar performance with System-I with lazy allreduce.

% In this section, we show the overall performance evaluation results of using System-I to train AlexNet and RestNet-50 on the ImageNet-1K dataset with 128 Pascal GPUs and 512 Volta GPUs. We first show the effectiveness of employed network optimization techniques (ring-based allreduce, mixed-precision training, computation/communication overlap, lazy allreduce and coarse-grained sparse communication) and their combinations. Then we show the System-I's overall training time, and compare it with other approaches.

% \renewcommand{\arraystretch}{1.15}
\begin{table*}[]
\centering
\caption{Compare AlexNet training with different approaches.}
\vspace{-5pt}
\label{Tab: alexperf}
\begin{tabular}{p{14em}C{5em}C{10em}C{9em}C{8em}C{8em}}
\toprule
& Batch Size & Processor & GPU Interconnect & Time & Top-1 Accuracy \\ \midrule
You et al. \cite{you2018imagenet} & 512   & DGX-1 station & NVLink & 6 hours 10 mins  & 58.8\%   \\
You et al. \cite{you2018imagenet} & 32K   & CPU x 1024 & - & 11 mins   & 58.6\%   \\
Jia et al. \cite{jia2018highly}  & 64K  & Pascal GPU x 512 & 100 Gbps & 5 mins & 58.8\%  \\
Jia et al. \cite{jia2018highly}  & 64K  & Pascal GPU x 1024 & 100 Gbps& 4 mins & 58.7\%   \\
{This Work (DenseCommu)}  & {64K} & {Volta GPU x 512} &{56 Gbps} &  {2.6 mins}  & {58.7\%} \\
{This Work (SparseCommu)}  & {64K} & {Volta GPU x 512} & {56 Gbps} &  {1.5 mins}  & {58.2\%}
\\ \bottomrule
\end{tabular}
\end{table*}

\begin{table*}[]
\centering
\caption{Compare ResNet-50 training with different approaches.}
\vspace{-5pt}
\label{Tab: resperf}
\begin{tabular}{p{14em}C{5em}C{10em}C{9em}C{8em}C{8em}}
\toprule
& Batch Size & Processor & GPU Interconnect & Time & Top-1 Accuracy \\ \midrule
Goyal et al. \cite{goyal2017accurate} & 8K  & Pascal GPU x 256 & 56 Gbps  & 1 hour   & 76.3\%   \\
Smith et al. \cite{smith2017don} & 16K  & Full TPU Pod & - & 30 mins  & 76.1\%   \\
Codreanu et al. \cite{codreanu2017scale} & 32K   & KNL x 1024 & - & 42 mins   & 75.3\%   \\
You et al. \cite{you2018imagenet} & 32K   & KNL x 2048 & - & 20 mins   & 75.4\%   \\
Akiba et al. \cite{akiba2017extremely} & 32K  & Pascal GPU x 1024 & 56 Gbps  & 15 mins  & 74.9\%   \\
Jia et al. \cite{jia2018highly}  & 64K  & Pascal GPU x 1024 & 100 Gbps & 8.7 mins & 76.2\%   \\
Jia et al. \cite{jia2018highly}  & 64K  & Pascal GPU x 2048 & 100 Gbps& 6.6 mins & 75.8\%   \\
Mikami et al. \cite{mikami2018imagenet} & 68K  & Volta GPU x 2176 & 200 Gbps  & 3.7 mins & 75.0\%  \\
This Work (DenseCommu)  & 64K & Volta GPU x 512 & 56 Gbps &  7.3 mins  & 75.3\%
\\ \bottomrule
\end{tabular}
\end{table*}

\subsection{End-to-End Training Time}

In this set of experiments, we measure the overall training time of AlexNet and ResNet-50 on the ImageNet-1K dataset. 
To maximize network performance of distributed DNN training, System-I enables ring-based allreduce, mixed-precision training, computation/communication overlap, lazy allreduce and coarse-grained sparse communication. The per-GPU batch size is 128 for both AlexNet and ResNet-50. The overall batch size of the training task is 64K with 512 GPUs.  To avoid losing model quality with large batch size training, System-I employs layer-wise adaptive rate scaling (LARS) \cite{you2018imagenet} algorithm, and make it work in conjunction with mixed-precision training. We adopt the linear scaling rule with warm-up scheme to adjust the learning rate. Also, System-I performs all data preprocessing tasks on GPUs to further improve system performance. We complete AlexNet training in 95 epochs, and complete ResNet-50 training in 90 epochs.

Jia et al. \cite{jia2018highly} could complete  ImageNet/AlexNet training  in 4 minutes on 1024 Pascal GPUs. As shown in Table \ref{Tab: alexperf}, we break this record and complete AlexNet training in 2.6 minutes using System-I without coarse-grained sparse communication. If we enable coarse-grained sparse communication with sparsity ratio 85\% (15\% of gradient chunks are selected for allreduce), System-I completes AlexNet training in 1.5 minutes. System-I provides shorter training time with less amount of GPUs and lower network bandwidth. Also, System-I achieves $>58\%$ top-1 accuracy in both cases.  

Table \ref{Tab: resperf} shows that System-I finishes ImageNet/ResNet-50 training in 7.3 minutes on 512 Volta GPUs without enabling sparse communication. Jia et al. \cite{jia2018highly} finish the training in 8.7 minutes with 1024 Pascal GPUs connected by 100 Gbps network. Compared to this work, System-I could also achieve shorter training time with less amount of GPUs and lower network bandwidth.  Since ResNet-50 training is computation intensive, adding more GPUs could achieve faster training speed. Mikami et al. \cite{mikami2018imagenet} achieve 1.97 speedup ratio than our work with 4x more GPUs and 4x higher bandwidth.

\section{Related Work}\label{sec:related_work}

In this section, we discuss a number of approaches to reduce communication overhead for distributed DNN training systems, which are designed based on PS or allreduce.

\textbf{High-Performance Network Fabrics.} 
S-Caffe \cite{awan2017s}, FireCaffe \cite{iandola2016firecaffe}, Malt \cite{li2015malt} and COTS-HPC \cite{coates2013deep} use  InfiniBand to improve the performance of distributed DNN training in HPC clusters. To further improve GPU-GPU communication, several collective communications library implementations, such as NCCL \cite{NCCL}, provide efficient CUDA-Aware support using techniques like GPUDirect RDMA \cite{awan2017s}, \cite{wang2011mvapich2}. GradientFlow also uses RDMA  to transmit gradients over InfiniBand for low latency.

\textbf{Reducing Communication Frequency.} 
DistBelief \cite{dean2012large} and Petuum \cite{Petuum} can reduce the communication frequency for PS-based distributed training.  Specifically, DistBelief allows worker nodes to pull latest parameters every $u$ iterations and push gradients every $v$ iterations, where $u$ might not be equal to $v$. In Petuum, worker nodes could use cached stale parameters for computing gradients, until the version of cached parameters is older than a threshold. In this way, DistBelief and Petuum could reduce communication frequency and network traffic for higher model training performance. However, it is still unclear that whether reducing communication frequency could still keep model quality when working with large batch size training.

\textbf{Sparse Communication.} Li-PS \cite{li2014communication} and deep gradient compression (DGC) \cite{lin2017deep} employ fine-grained sparse communication to reduce network traffic for distributed DNN training. Li-PS uses the PS architecture, and DGC proposes sparse allreduce. In both systems, a GPU transmits a small part of important gradients instead of all gradients. For example, only gradients greater than a threshold can be selected for transmission. Compared to DGC and Li-PS, GradientFlow leverages coarse-grained sparse communication to reduce network traffic with high bandwidth utilization by selecting important gradient chunks for allreduce. With this design, GradientFlow could still use dense communication librarie like NCCL for sparse communication.

\textbf{Communication/Computation Overlapping.}  Petuum \cite{Petuum} and B\"osen \cite{wei2015managed} could  overlap communication with computation for PS-based distributed DNN training.  These two systems  use the stale-synchronous parallel synchronization model to  overlap the communication of previous iterations with the computation of current iteration. Such communication/computation overlap mechanism  may require more iterations to complete model training or could reduce model quality, since GPU worker nodes cannot use latest parameters to compute gradients. Compared to Petuum and B\"osen, GradientFlow supports to overlap the communication of top DNN layers with the computation of bottom layers in a single iteration, and would not reduce model quality.

\textbf{Model Compression. } CNTK \cite{seide2014parallelizability} and Poseidon \cite{zhang2015poseidon} support to use compression techniques to reduce the size of transmitted gradients for PS. Specifically,  Poseidon uses  sufficient factors to represent dense fully connected layers of DNNs. CNTK represents gradients as 1-bit values for speech DNNs with some  negative impacts on model accuracy. GradientFlow scales  mixed-precision training to a distributed setting, which reduces network traffic by transmitting fewer bits for the same number of gradients.

\textbf{Constant Communication Cost. } Ring-based allreduce \cite{Baidu-AllReduce} and BytePS \cite{BytePS} could achieve high system scalability  with constant communication cost. When using ring-based allreduce to exchange $K$ bytes of data with $N$ GPUs, each machine  sends and receives $2(N-1)K/N$ bytes of data, which is independent with $N$ when $N$ is large. Horovod \cite{sergeev2018horovod} proposes gradient fusion scheme to address small message transmission problem for  ring-based allreduce in large-scale clusters.  In PS-based systems with $N$ GPU worker nodes and $M$ CPU server nodes, each GPU worker node pushes and pulls $K$ bytes of data in an iteration, and each CPU server node sends and receives $KN/M$ bytes of data. BytePS addresses server node communication bottleneck by setting up more server nodes, for example $N=M$. In this case, BytePS has half constant communication cost than ring-based allreduce. However, BytePS requires heterogeneousness cluster to provide additional CPU machines. GradientFlow is designed for distributed DNN training on homogeneous clusters: each machine is installed with same type of GPU, CPU and network device. We may waste half of GPU computation power when using BytePS in a high performance homogeneous cluster.

\section{Conclusion}

 In this paper, we propose a communication backend named GradientFlow to improve network performance for distributed DNN training. Our first contribution is showing the performance gain of recently proposed ring-based allreduce, mixed-precision training, and computation/communication overlap. Our results show that a real system with aforementioned approaches still has huge performance gap with the ideal system. To further reduce network cost, we propose lazy allreduce to improve network throughput by fusing multiple communication operations into a single one, and design coarse-graining sparse communication to reduce network traffic by only transmitting important gradient chunks. The results show that GradientFlow could significantly reduce the communication overhead of distributed DNN training. When training AlexNet and ResNet-50 on the ImageNet dataset using 512 GPUs, our work could  achieve 410.2 and 434.1 speedup ratio respectively. 
 In the future, we would continue to optimize the communication performance of distributed DNN training on modern hardwares, such as NVLink, 100/200Gbps RDMA network with in-network computation capability and protocol like SHArP \cite{graham2016scalable}. We would also consider network performance optimization when training DNNs over heterogeneous HPC clusters with different intra-server throughputs and inter-server network bandwidth.

\section*{Acknowledgement}

We gratefully acknowledge contributions from our colleagues within SenseTime Research and our collaborators from Cloud Application and Platform Lab, Nanyang Technological University Singapore.

\bibliographystyle{IEEEtran}
\bibliography{main}

\balance

\end{document}